\documentclass[11pt]{article}
\usepackage{graphicx}
\usepackage{color}
\usepackage{hyperref}

\newtheorem{thm}{Theorem}

\newtheorem{tab}{Table}
\newtheorem{fig}{Figure}

\def\leurre{\noindent\leftskip0pt\small\baselineskip 10pt}

\def\ligne#1{\hbox to \hsize{#1}}
\setbox110=\vtop{
\normalsize\baselineskip 13pt
\ligne{\hfill Maurice \sc Margenstern\hfill}
\ligne{\hfill
Laboratoire d'Informatique Th\'eorique et Appliqu\'ee, EA 3097,\hfill}
\ligne{\hfill Universit\'e de Lorraine,\hfill}
\ligne{\hfill Campus du Saulcy,\hfill}
\ligne{\hfill 57045 Metz Cedex, France,\hfill}
\ligne{\hfill {{\it email:} \tt maurice.margenstern@univ-lorraine.fr}\hfill}
\ligne{\hfill {{\it email:} \tt margenstern@gmail.com}\hfill}
}
\title{A weakly universal cellular automaton with 2 states on the
tiling $\{11,3\}$
\box110
}
\begin{document}
\maketitle

\begin{abstract}
In this paper, we construct a weakly universal cellular automaton with two states only 
on the tiling $\{11,3\}$. The cellular automaton is rotation invariant and it is a true planar one.
\end{abstract}

\section{Introduction}

The paper makes use of the improvements introduced in the paper~\cite{mm73uhca3st}. As most of the
author's papers in this topic, this paper makes use of the railway model, 
see \cite{stewart,mmbook3}. We just remind the reader that the circuit simulates a register
machine instead of a Turing machine as in \cite{stewart}. The Euclidean circuit consists in 
tracks which are either straight segments or quarters of a circle. The circuit allows crossings
and it makes use of three kinds of switches: the fixed one, the flip-flop and the memory switch.
A unique locomotive runs over the circuit and the evolution of the positions of the switches 
in time allows us to simulate a computation.
Although initially devised for the Euclidean plane, this model can be implemented in the 
tessellations of the hyperbolic plane, especially in the tessellations $\{p,4\}$ and
$\{p$$+$$2,3\}$ which are spanned by a tree which, in each case allows a rather easy way to 
implement
horizontals and verticals.

The first implementation, in 2002, was performed in the pentagrid, the tiling $\{5,4\}$ of the 
hyperbolic plane, see~\cite{fhmmTCS}. It required 22~states. The number of states was lowered 
down to~9{} in the same tiling in 2008, see~\cite{mmsyPPL}. At the same time, the
circuit was implemented in the heptagrid, the tiling $\{7,3\}$ of the hyperbolic plane,
see~\cite{mmsyENTCS}, requiring 6 states. Both papers of 2008 gave the same implementation. 
The smaller number of states is compensated by a higher number of neighbours. 
A bit later, I reduced the number of states down to~4{} in the heptagrid by a slight
change in the locomotive: replacing the previously green front cell by a blue one,
the same colour as that of the milestones. This smaller number of states in the heptagrid
compared to the pentagrid is not surprising. It was again observed a bit later in 2012 with 
the implementation in
the tiling $\{13,3\}$ of a weakly universal rotation invariant and planar cellular automaton
with two states only, see~\cite{mmarXiv2st,mmbook3}. This latter implementation introduces
many features which allowed to lower the number of states. The first idea was to mark the
tracks by milestones instead of assigning a specific colour to the tracks. Then, the idea was
to allow one-way tracks only. This implies a strong change in the switches. The fixed one is 
simplified to a passive switch only: the switch is only needed when the locomotive comes from one
of the two tracks which join into a single one at the switch. The flip-flop was already
a one-way switch in its original implementations as it must be crossed actively only. The constraint
of one-way tracks implies to split the memory switch into two switches: an active one and a passive
one. The working of the switch makes it necessary to connect the active switch to the passive
one: when the locomotive comes from the non-selected track in the passive switch, the selection
changes there and it also must change in the active switch. 

These new features where enough to reduce the number of states down to two of them in the
dodecagrid, the tiling $\{5,3,4\}$ of the hyperbolic $3D$-space. The space allows us to 
get rid of the crossings, which is not at all the case in the plane. Moreover, the introduction
of one-way tracks makes the number of crossings to seriously increase: a previous two-way crossing
has to be implemented by four one-way crossings. The implementation in $\{13,3\}$ introduced a
new feature coming from roadway traffic, the round-about. It also reduced the locomotive 
to a single cell, making it closer to a particle. Trying to implement these new ideas in
the heptagrid, I recently obtained a weakly universal cellular automaton with three states only,
see~\cite{mm73uhca3st}. Now, in this paper, I had to improve the implementation of the
one-way switches. In particular, introducing patterns used in asynchronous cellular automata,
I could organize the implementation of the flip-flop and the active-memory switch
in a similar way, using two new and simpler components, the fork and the selector. The difference 
in the switches is now
given by the assembling of these new components, a bit different in these two switches.
I thought it could be useful to implement these new elements in order to obtain a cellular
automaton with two states in a grid $\{p,3\}$ with \hbox{$p<13$}. I could do that for 
\hbox{$p=11$} by replacing the passive memory switch by a combination of the fork with a 
new structure, the controller, a bit simpler than the passive memory switch itself.

We make all of this more precise in Section~\ref{scenar}.

Before turning to Section~\ref{scenar}, let us explain how we shall illustrate
our implementation. The tiling $\{11,3\}$ in the Poincar\'e's disc model is illustrated by 
the leftmost picture of Figure~\ref{til11_3}. 

\vtop{
\vspace{5pt}
\ligne{\hfill
\includegraphics[scale=0.56]{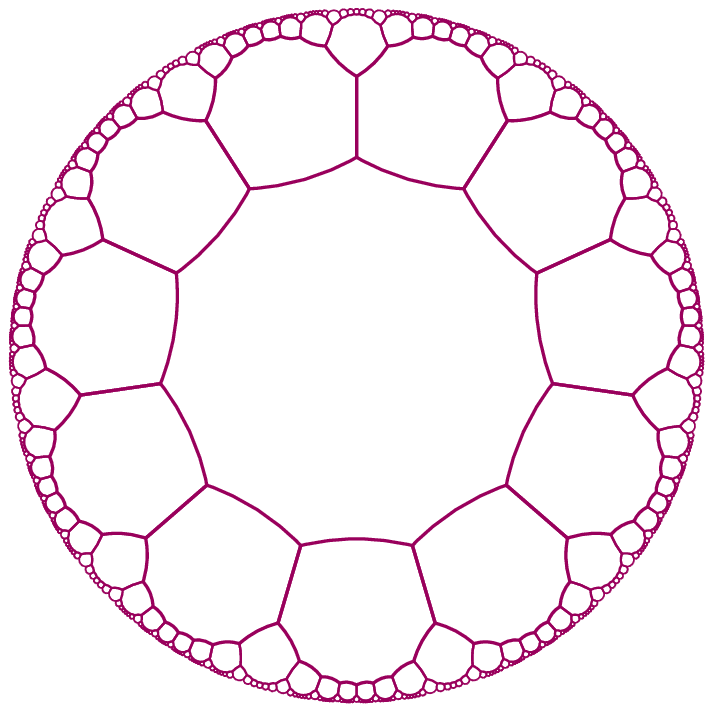}
\hfill
\includegraphics[scale=0.47]{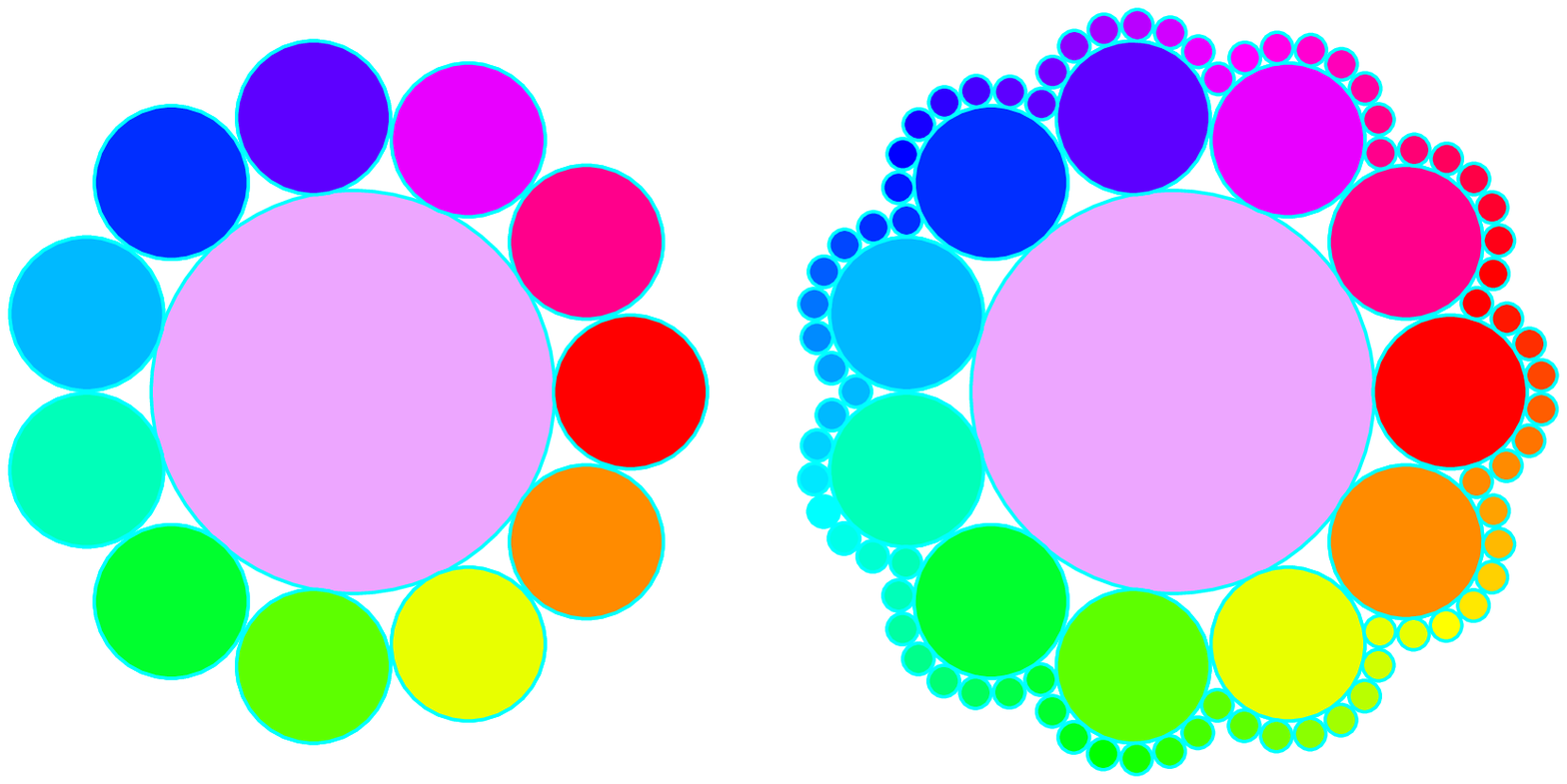}
\hfill}
\vspace{-5pt}
\begin{fig}\label{til11_3}
\leurre
To left: the tiling $\{11,3\}$ in the Poincar\'e's model. To right:
another representation of a cell in the tiling $\{11,3\}$ and the same one together
with its neighbours.
\end{fig}
}


We do not see very much in this representation, so that we shall replace it by the two
other illustrations given in Figure~\ref{til11_3}. As will be seen in the further 
illustrations, this representation 
allows us to give it some flexibility which will improve the readability of the figures.
     
\section{The scenario}
\label{scenar}

   In this section, we precisely describe the implementation of the railway circuit.
Here, we admit that it is possible to devise such a circuit that the motion of the locomotive
simulates the computation of a register machine thanks to the positions of all switches of the
circuit. This is explained with all details in~\cite{mmbook3}. 

Let us remind the reader that the systematic one-way track organization of the circuit leads
to a more complex representation of the memory switch. On the left-hand side of
Figure~\ref{memoswitch_0} we have a sketchy representation of the two-way memory switch.
We remind the reader that the two-way switch may be crossed either passively or actively.
On the right-hand side, we have the one-way switch. It consists in two one-way half-switches:
an active half-switch on the left-hand side and a passive one on the right-hand side. A connection
goes from the passive half-switch to the active half-switch: this is needed when the selection
has to be changed. It first changes at the passive part which detects that the locomotive came
from the non-selected track, and the necessity to change the selection is passed to the
active part through the orange path of Figure~\ref{memoswitch_0}.

\vtop{
\vspace{-15pt}
\ligne{\hfill
\includegraphics[scale=0.7]{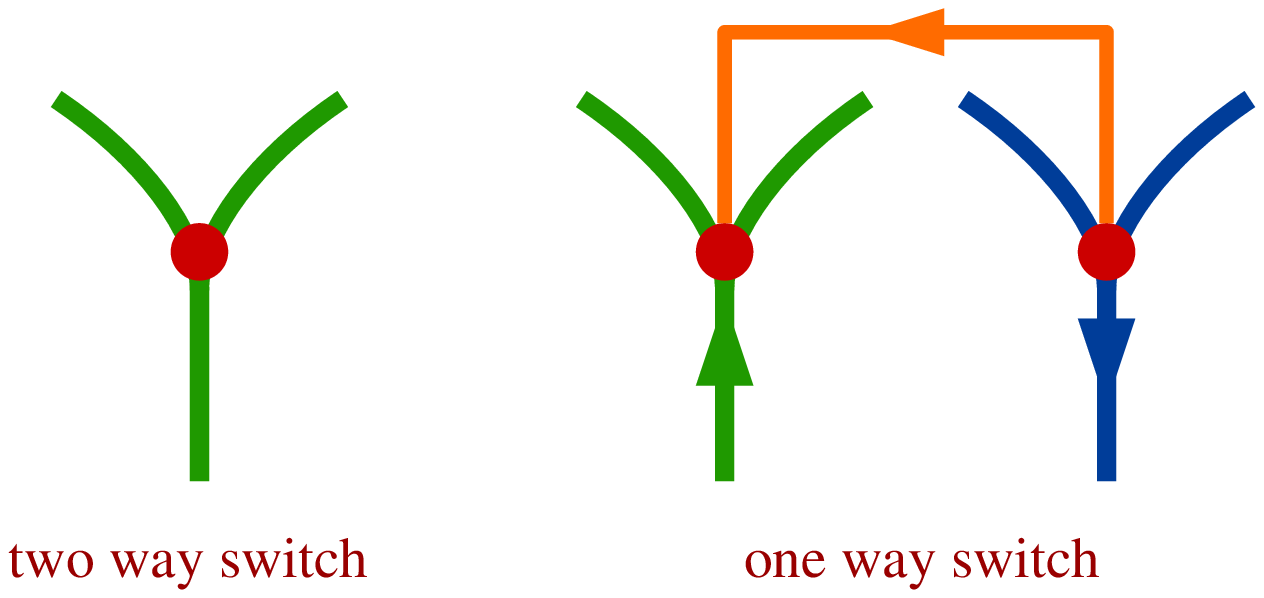}
\hfill}
\vspace{-10pt}
\begin{fig}\label{memoswitch_0}
\leurre
Comparison between the two-way and the one-way representation of the memory switch.
\end{fig}
}

In this section, we precisely
explain the new features in Sub-section~\ref{genpatt} and how they are implemented in
Sub-section~\ref{implements}.

\subsection{General patterns}
\label{genpatt}

   Let us first list the elements we shall study in Sub-section~\ref{implements}. With each name of
an element, we sketchy describe what the element is expected to perform.

   First, the {\bf elements of the tracks}: each cell of the track is marked with appropriate 
milestones. Such an element may be crossed either by a single locomotive or by two locomotives
running together, contiguously. We shall see the elements in Sub-section~\ref{implements}. 

Then, the second structure we need is the {\bf fixed switch}. As mentioned in the introduction,
for one-way tracks it is a passive structure only. It gathers two tracks which are melted into
a single one after the cell at which the two tracks arrive.

   Next, we have two patterns involved by the round-about: the {\bf duplicator} and the 
{\bf selector}.

   The duplicator has two points of connection with the tracks: an {\bf entrance}
and an {\bf exit}. The locomotive arrives through the entrance. Two locomotives
leave the duplicator through the {\bf exit}. 

    The selector has three points of connection with the tracks: 
an {\bf entrance} and two {\bf exits}. Each exit correspond to the number of locomotives
entering the selector. If a single locomotive enters, it leaves the selector through {\bf exit~1}
which is connected with a track leaving the round-about. If two contiguous locomotives enter
the selector, one of them leaves the selector through {\bf exit~2}. Exit~2 is attached to a piece
of tracks leading to another selector. 

\vtop{
\ligne{\hfill\hskip-40pt
\includegraphics[scale=0.7]{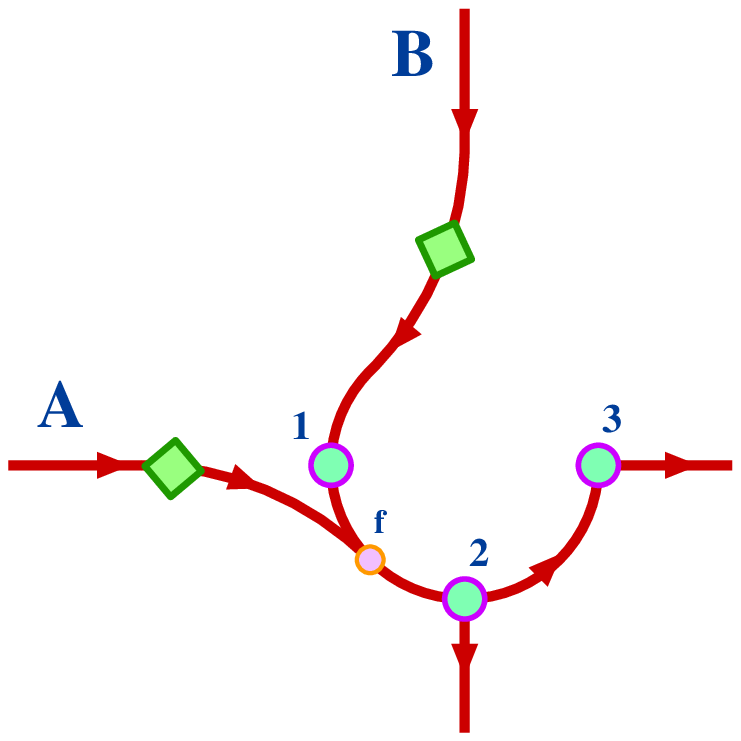}\hfill}
\vspace{-15pt}
\begin{fig}\label{roundabout}
\leurre
The round-about. The arrows indicate the path used by the locomotive. 
\end{fig}
}

   This explains how a {\bf round-about} works. This is illustrated 
by Figure~\ref{roundabout}. In the figure, the duplicator is illustrated by a green rhomboid 
pattern. The selectors are illustrated by a disc. A round-bout assembles two duplicators, three
selectors and a fixed switch, denoted by~{\tt f} in the figure. The locomotive crosses one disc and
leaves the round-about at the second one. When it arrives through~{\bf A}, {\bf B}, the locomotive
is transformed into two contiguous locomotives after crossing the duplicator. When it meets
the first selector, one locomotive is killed while the second one goes on its way on the 
round-about, towards the second selector. At the second selector, 3, 2 respectively, it leaves the
round-about to continue its way on the same tracks.

  Next, we consider the {\bf controller}, a structure which is used by both the flip-flop
and the active memory switch. The controller has two states~: {\tt accept} or {\tt reject}. If on 
{\tt accept}, the locomotive entering the controller is accepted and it is allowed to cross
it in order to go on its way on the tracks. If the controller is on {\tt reject}, then the
locomotive is not accepted: it does not leave the pattern, it is killed there.

   Next we consider the {\bf fork}. It is different from the duplicator: here too, one locomotive
enters and two ones exit, but the two exiting locomotives are on different tracks. The fork has
an {\bf entrance} and two {\bf exits}. When the locomotive has crossed the fork, there is one 
locomotive on each track leaving the pattern.

\vtop{
\ligne{\hfill\includegraphics[scale=0.6]{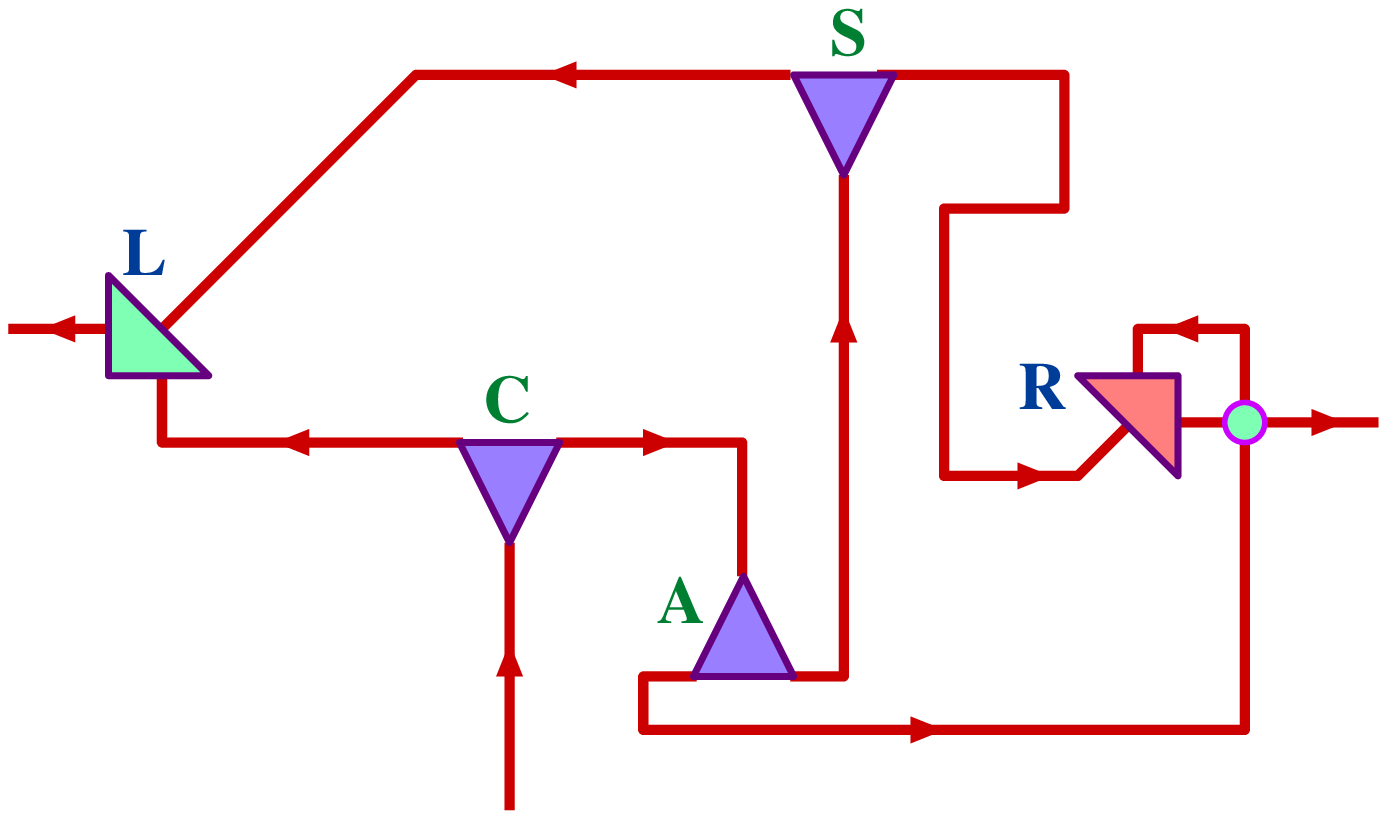}
\hfill}
\vspace{-15pt}
\begin{fig}\label{flipflop}
\leurre
Association of forks and controllers in order to constitute a flip-flop. 
\end{fig}
}

   Both structures, the controller and the fork are used to implement a flip-flop
and an active memory switch. Figure~\ref{flipflop} shows how to combine three forks and two
controllers in order to obtain a flip-flop.

\vtop{
\vspace{-5pt}
\ligne{\hfill
\includegraphics[scale=0.7]{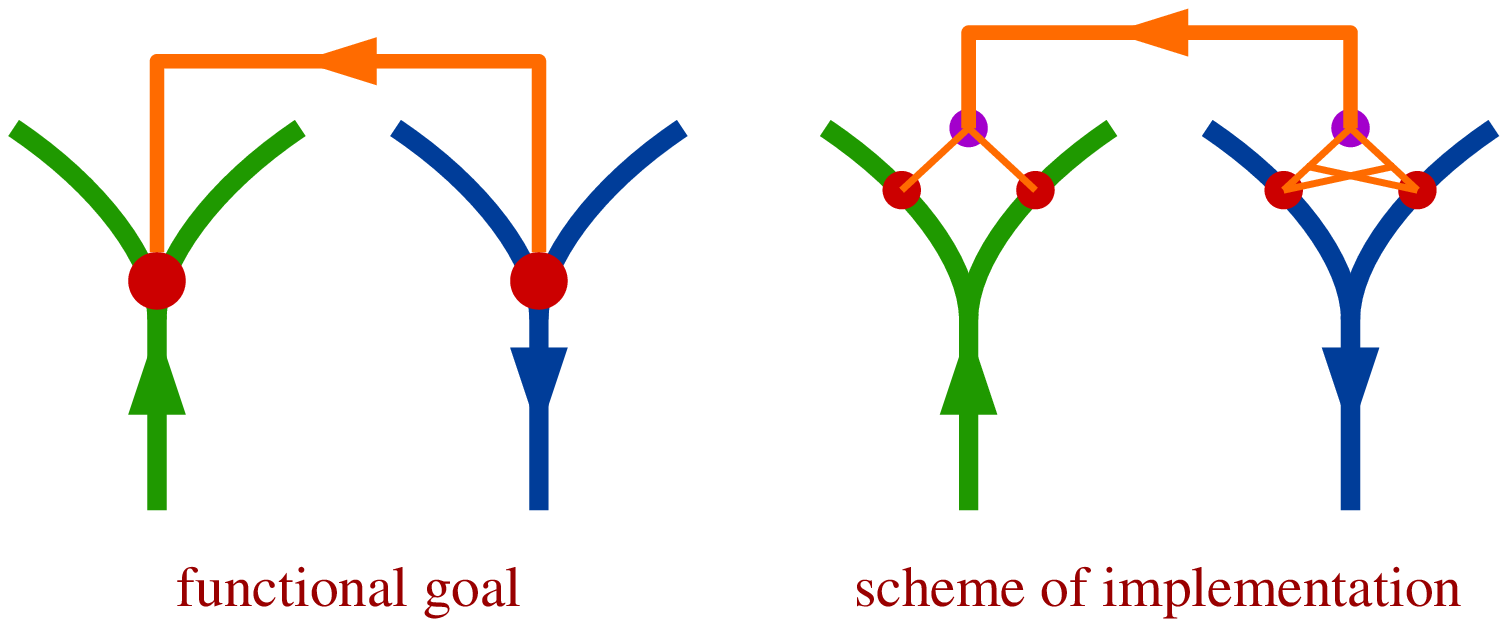}
\hfill}
\vspace{-10pt}
\begin{fig}\label{memoswitch_1}
\leurre
To left: the functional goal of the new implementation of the memory switch with one-way
tracks. To right: the scheme of implementation.
\end{fig}
}

Note that this is connected with what we have indicated in 
Figure~\ref{memoswitch_0}. Now it is time to make this idea more precise. 
Figure~\ref{memoswitch_1} gives the general scheme of implementation of the memory switch
under the constraint of one-way tracks.

Now we turn to a more exact implementation of the scheme illustrated by the right-hand side
part of Figure~\ref{memoswitch_1}. First,
Figure~\ref{memoactive} shows how to combine the same elements in order to obtain 
an active memory switch, following the scheme of Figure~\ref{memoswitch_1}. 

\vtop{
\vspace{-15pt}
\ligne{\hskip 0pt\includegraphics[scale=0.6]{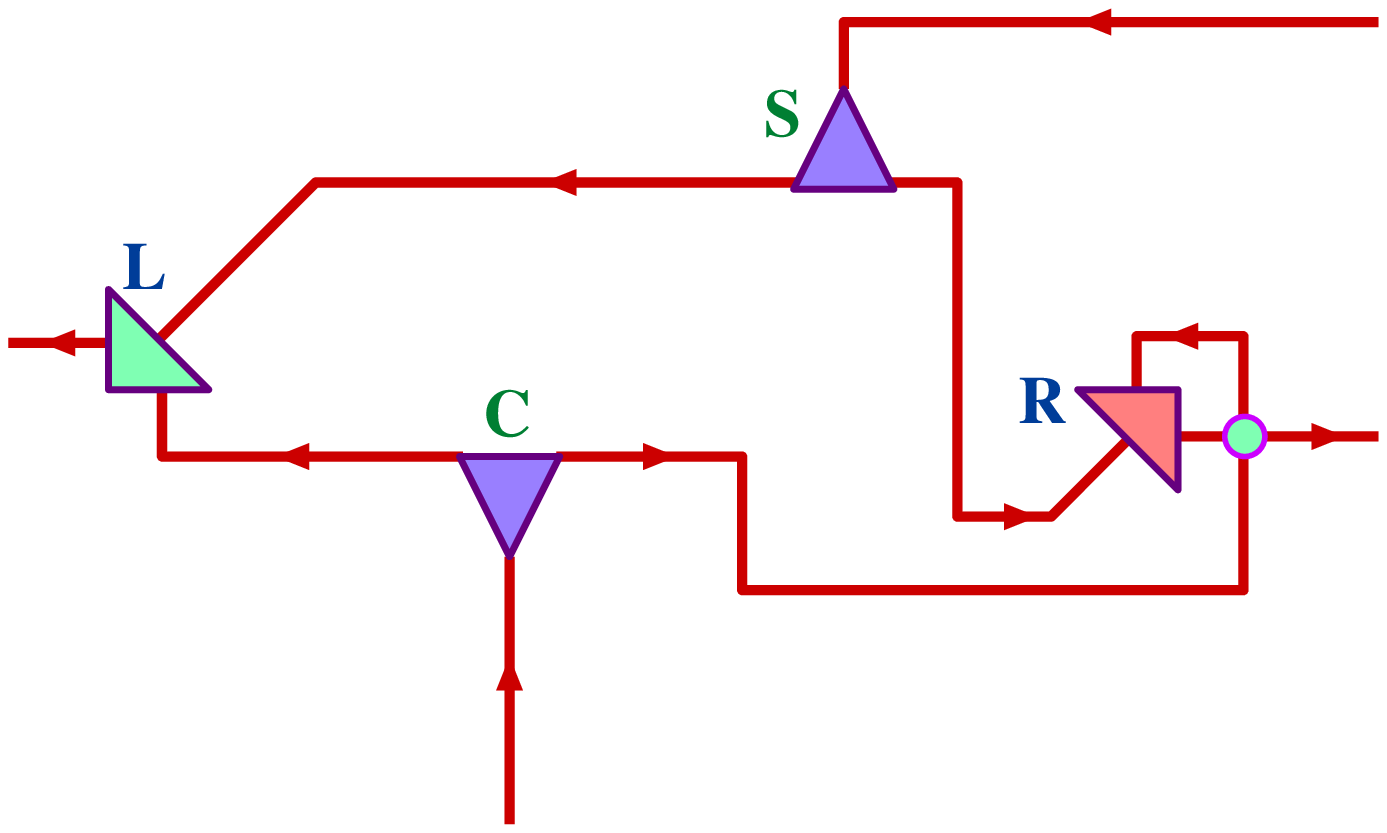}
\hfill}
\vspace{-15pt}
\begin{fig}\label{memoactive}
\leurre
Association of forks and controllers in order to constitute an active memory switch. 
\end{fig}
}

\vtop{
\vspace{-5pt}
\ligne{\hskip 0pt\includegraphics[scale=0.6]{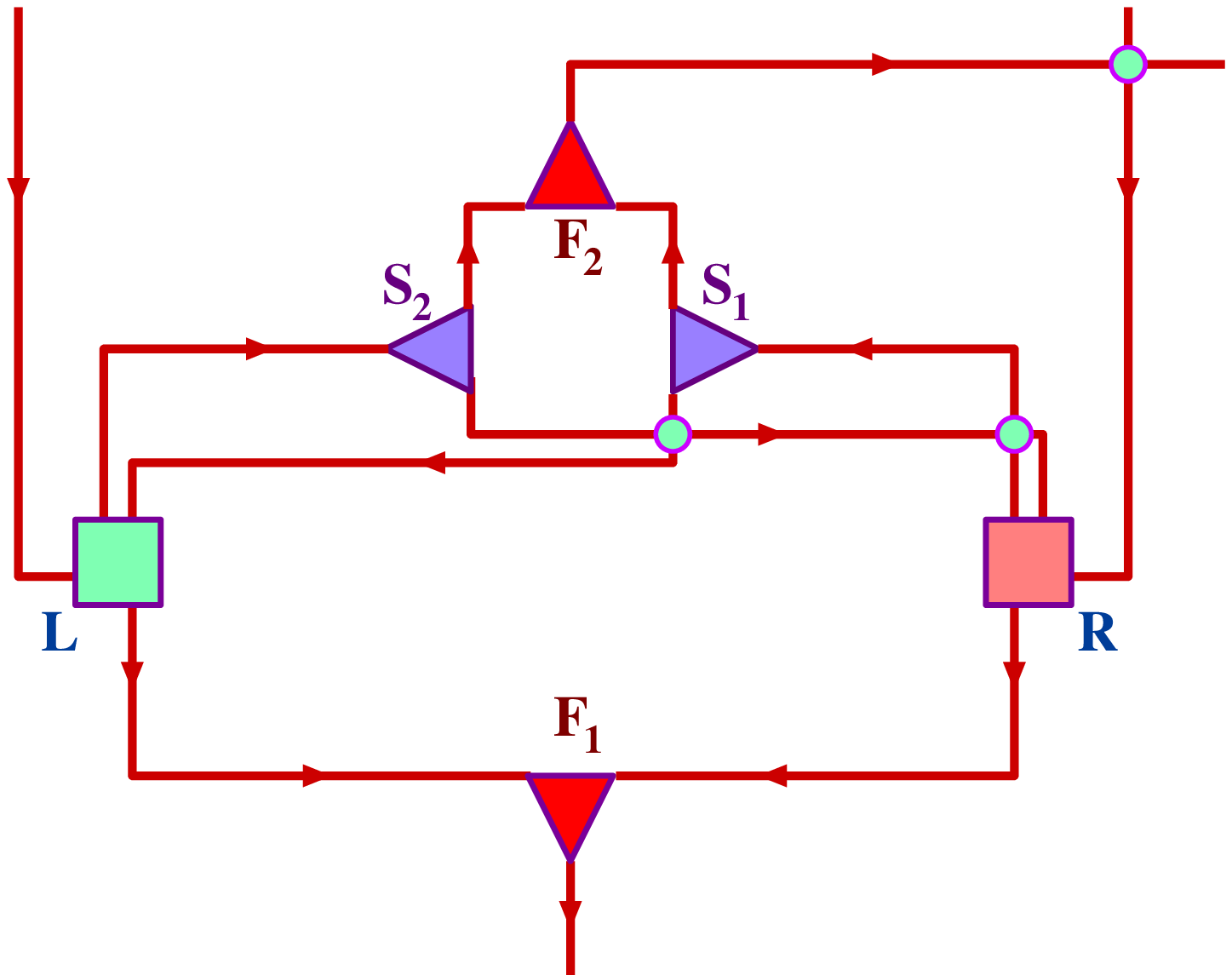}
\hfill}
\vspace{-10pt}
\begin{fig}\label{memopassive}
\leurre
Association of forks and new controllers in order to constitute a passive memory switch. 
\end{fig}
}

   In order to implement the scheme proposed by Figure~\ref{memoswitch_1}, we have to
introduce a
a new element. 
Note that the idea is to align the the presentation of the passive
memory switch with that of the active one by splitting a unique control at the switch
into two independent controls on each branch of the tracks going to the switch.
Now, due to the basic difference of working of the switches, we cannot use the same controller 
as in the active memory switch. The reason is that here, the controller must not stop the 
locomotive which crosses its structure. It has only to react to such a passage in case it is 
not the selected one.
This means that the controller performs a double function, but not at the same time.
When the controller is on the presently non-selected track, it changes the selection and, at 
the same time, it sends a signal to the active memory switch in order to change the selection
there too. 

    Figure~\ref{memopassive} illustrates how the fixed switch, the fork and the new controller
have to be associated in order to implement a passive memory switch, following the
scheme of implementation given in Figure~\ref{memoswitch_1}. Note that the track leaving
the switch {\tt F}$_{\hbox{\tt 2}}$ in Figure~\ref{memopassive} is the track arriving to the
fork~{\tt S} in Figure~\ref{memoactive}.

\subsection{Detailed implementations}
\label{implements}

   Now, let us turn to the exact implementation of the patterns described in 
Sub-section~\ref{genpatt}. We shall intensively use the representation introduced by
Figure~\ref{til11_3}. Now, as can be noticed already from Figure~\ref{voie11_1}, we shall
not always represent the neighbours of the same cell by circles or coloured discs of the same
size. The size of a neighbour will mainly be dictated by its role in the considered configuration.
In particular, the necessity to represent the neighbourhood of this neighbour will also play
a role. 

We shall successively study the elements of the tracks, 
the duplicating structure, the selector, then the fork, the controller for the active switches
and the controlling structure of the passive memory switch. We shall study 
{\bf idle configurations} only, {\it i.e.} configurations when the locomotive is not in the 
neighbourhood of the structure. We shall illustrate the motion of the locomotive through the
structures when we shall establish the rules, see Section~\ref{rules}.

\subsubsection*{The tracks}
\label{thetracks}

   The tracks are the first object we have to implement. We must not neglect this point. First
of all, without tracks, the information of what happens at some switch will for ever remain in
the switch, which is of no use for the computation. Second, the tracks certainly constitute the
biggest part of the circuit in term of quantity of involved elements. Transported into the 
hyperbolic plane, the horizontal parts of the tracks represented in Figures~\ref{flipflop}
to~\ref{memopassive} require a huge amount of cells.

Using Figure~\ref{elem_voie} of the present sub-subsection,
we can see on 
Figure~\ref{voie11_1} how we can implement a track going from a cell to any other one.
Note that in this figure,
we make use of the elements of the lower row in Figure~\ref{elem_voie}.
In principle, we could make use of these elements only. Indeed, considering two fixed 
non-neighbouring cells~$P$ and~$Q$. Consider a shortest path from the tile supporting~$P$
to that which supports~$Q$. Figure~\ref{elem_voie} how to organize a track joining~$P$ 
to~$Q$, using these elements only. Now, we use also the elements of the first row
of Figure~\ref{elem_voie} in order to make the implementation of the other structures
easier. In particular this will be used in the case of the round-about.

\vtop{
\vspace{-10pt}
\ligne{\hfill
\includegraphics[scale=0.4]{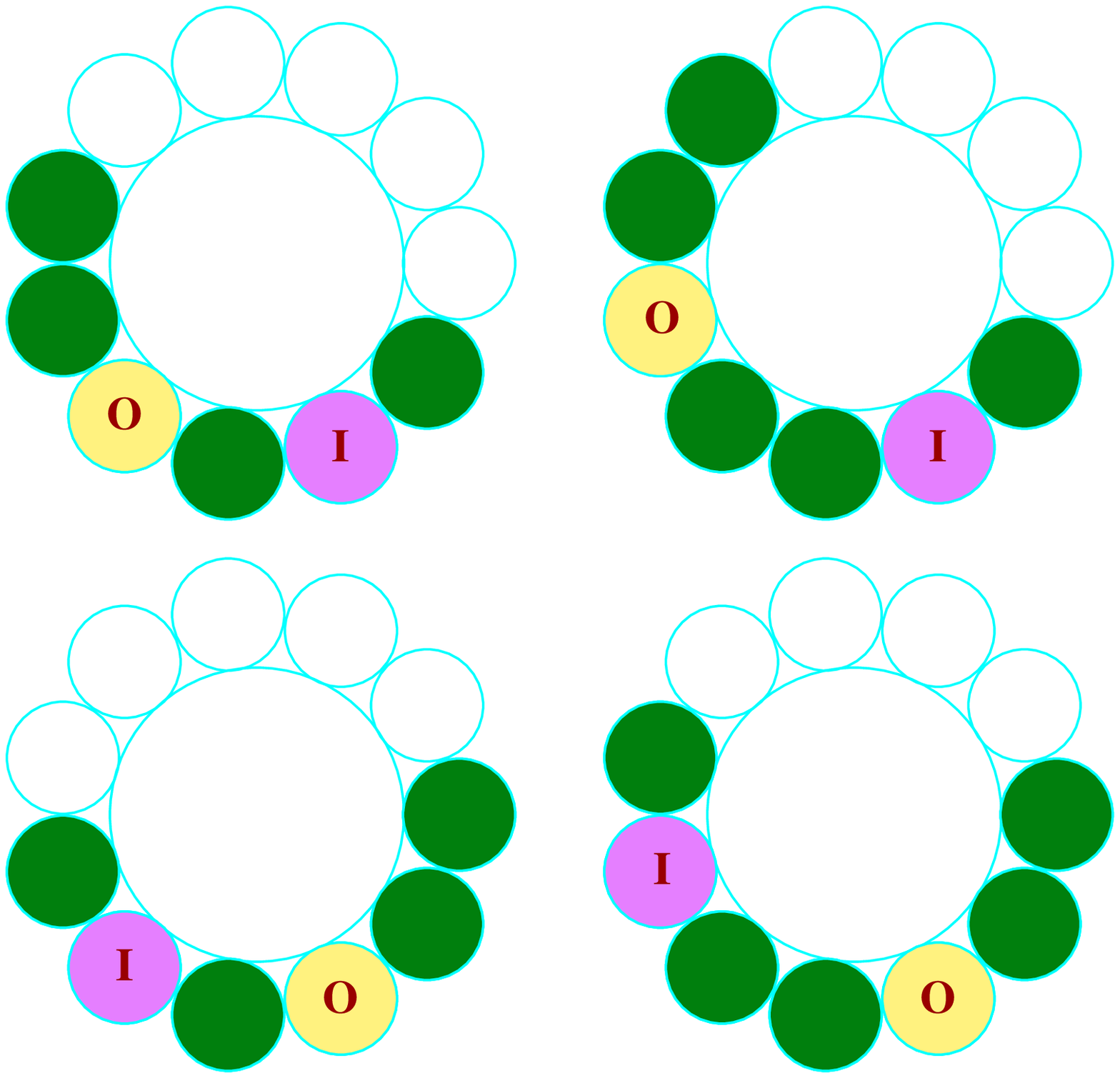}
\hfill}
\vspace{-10pt}
\begin{fig}\label{elem_voie}
\leurre
The four possible elementary elements of the tracks.
\end{fig}
}

\vtop{
\vspace{-20pt}
\ligne{\hfill
\includegraphics[scale=0.55]{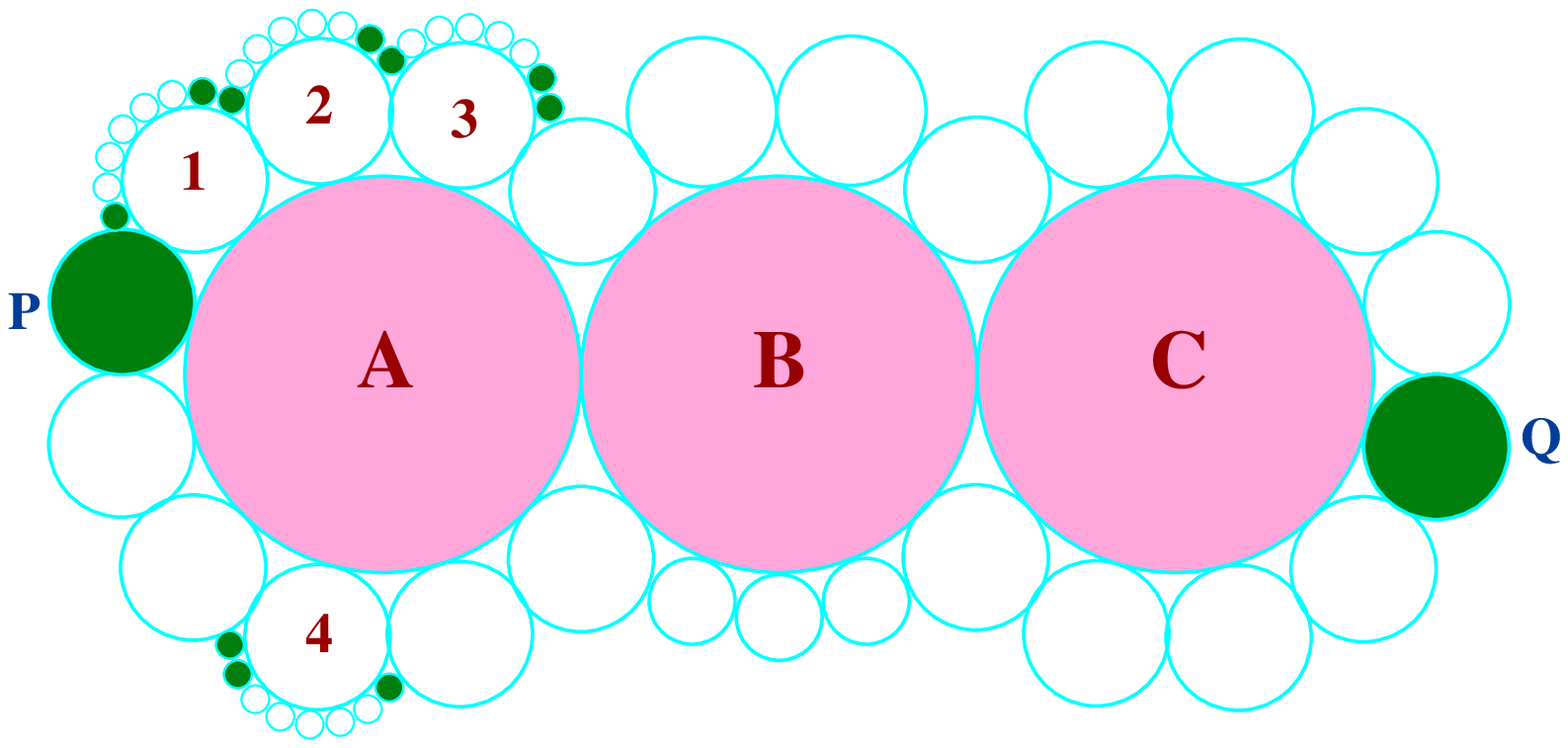}
\hfill}
\vspace{-25pt}
\begin{fig}\label{voie11_1}
\leurre
The use of elements of the tracks in order to define a track going from the cell~{\tt P} to
the cell~{\tt Q}.
\end{fig}
}

Note that in Figure~\ref{elem_voie},
the first column deals with cells of the track which turn around a fixed
black cell: call it the {\bf pivot} of the cell. But, as we can see on Figure~\ref{voie11_1},
it is needed to consider the case when the tracks go from one pivot to another one. This case
is dealt with by the second column of Figure~\ref{elem_voie}. For such cells, we say that it
is in between two pivots.


    Next, Figure~\ref{fix11} illustrates the implentation of the passive fixed switch.
Note the particular configuration of the entrances to the switch. They are elements of the track
but their working is a bit different as will appear in Section~\ref{rules}.

\vtop{
\vspace{-20pt}
\ligne{\hfill
\includegraphics[scale=0.55]{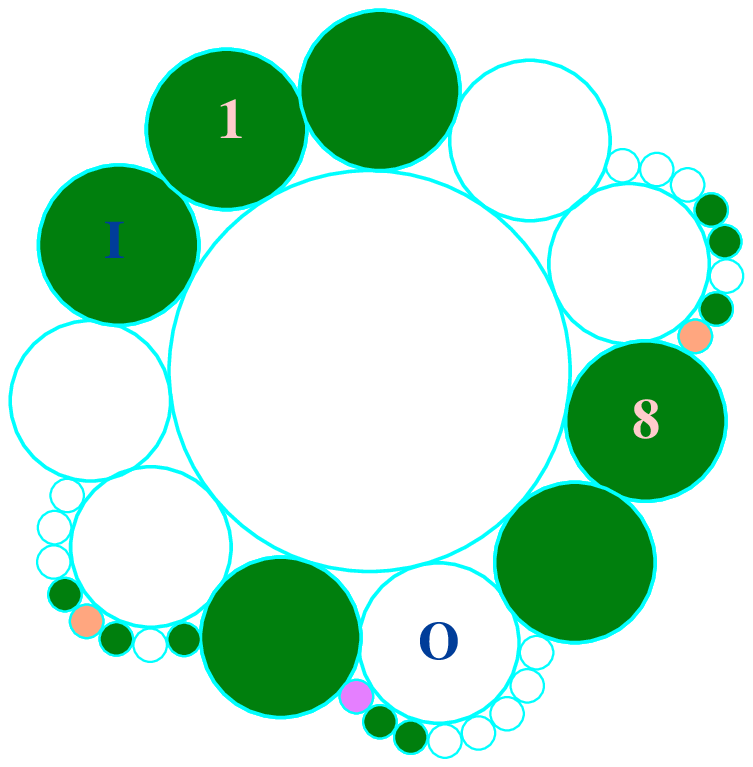}
\hfill}
\vspace{-10pt}
\begin{fig}\label{fix11}
\leurre
The idle configuration of the fixed switch.
\end{fig}
}

\subsubsection*{The round about}

   In this sub-subsection, we successively study the duplicator and the selector.

The idle configuration of the duplicator is illustrated by Figure~\ref{dupliq11}.
From Sub-section~\ref{genpatt}, we know that a single locomotive enters the structure
and that two contiguous ones leave it. 

\vtop{
\vspace{-10pt}
\ligne{\hfill
\includegraphics[scale=0.55]{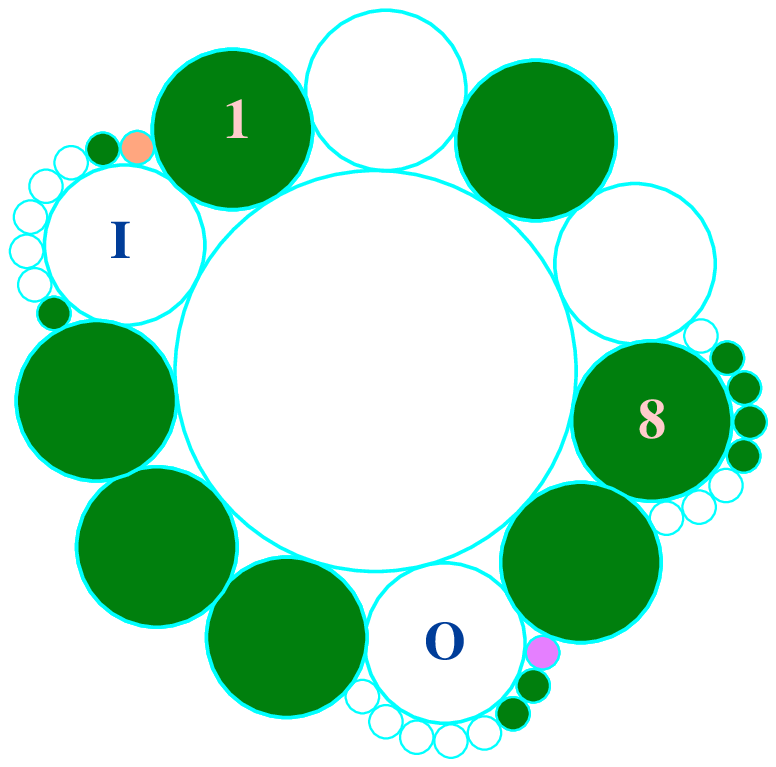}
\hfill}
\vspace{-10pt}
\begin{fig}\label{dupliq11}
\leurre
The idle configuration of the duplicator in a round-about.
\end{fig}
}

The single locomotive enters through the cell marked by~$I$
in the figure, while the two locomotives successively leave through the cell marked by~$O$.
The presence of the locomotive makes cell~8 flash by turning to white and then, at the next time,
by turning back to black. When cell~8 is white, the main cell remains to be black which creates
the needed second locomotive while the first one leaves the main cell. After that, cell~8 r
eturns to black so that the second locomotive also leave the duplicator.

   After the duplicator, we now look at the selector. It is a more complex structure. It has
to count how many locomotives arrive at the device and then, depending on whether it is the case
of one or two locomotives, it reacts in different ways. 

\vtop{
\vspace{-5pt}
\ligne{\hfill\hskip 15pt
\includegraphics[scale=0.5]{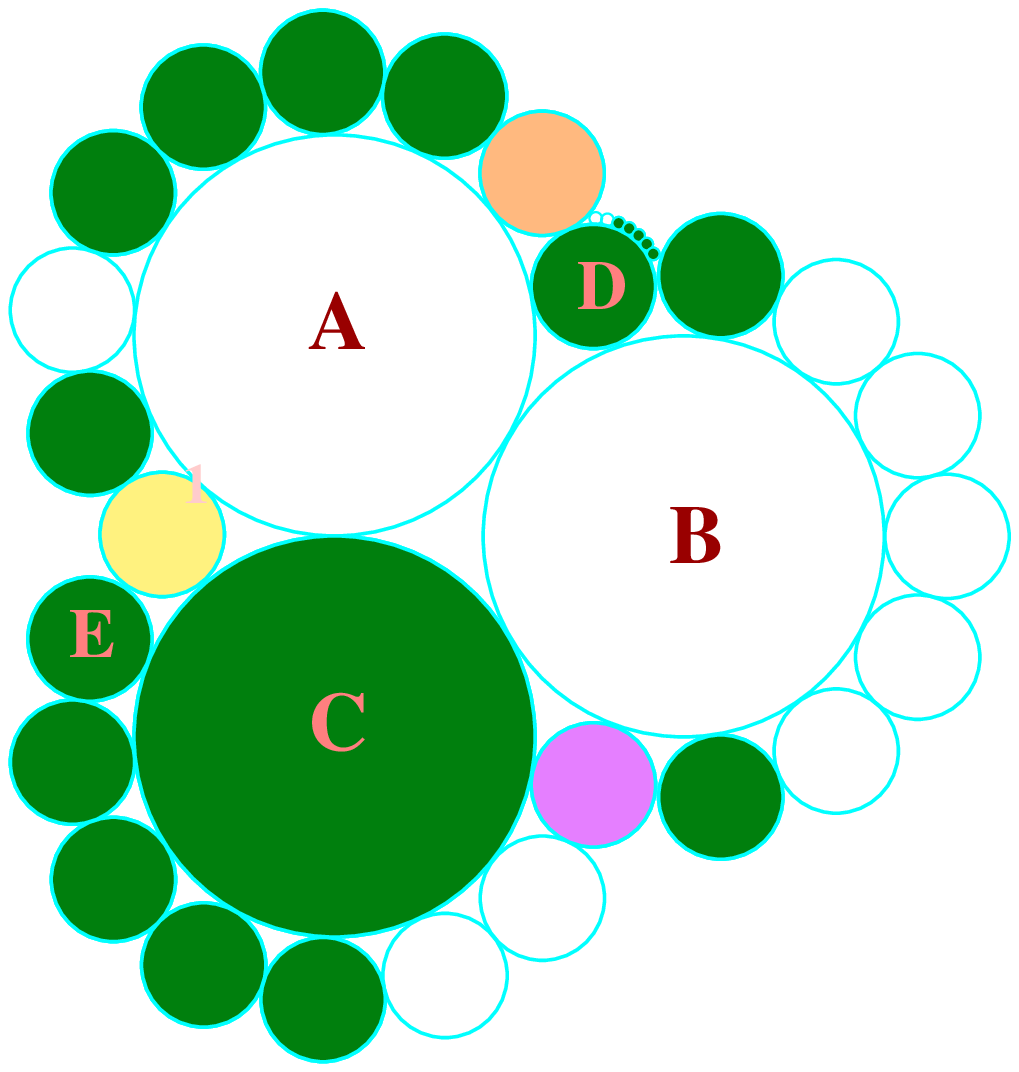}
\hfill}
\vspace{-10pt}
\begin{fig}\label{select11}
\leurre
The idle configuration of the selector used by a round-about.
\end{fig}
}

The locomotive arrive to the structure
by a the common neighbour of~$B$ and~$C$ at the bottom of~$B$ in Figure~\ref{select11}.
Then, the locomotive arrives to~$A$. There, the locomotive is duplicated on the two
exits from~$A$~: the exit which goes along~$E$ and the one which goes along~$D$, see the figure.
Two cells can see both~$A$ and~$B$: the cells~$C$ and~$D$. This allows the structure to count how
many locomotives arrive at it. We know that this number is one or two so that, writing
the state of~$A$ and then that of~$B$, the configuration seen of ~$A$$B$ seen from~$C$ and~$D$
is {\footnotesize\tt BW} in the case of one locomotive and {\footnotesize\tt BB} in the case
of two of them. If two locomotive arrive, one is killed, this means that $A$ is black for one
time only and $D$ turns to white while $C$ remains black. The effect of this action is that the
locomotive which arrives close to~$D$ is killed while that which was created at the common 
neighbour between~$C$ and~$E$ goes on along the track which will lead it to the next
selector of the round-about. When a single locomotive arrives, so that the
configuration seen from~$C$ and~$D$ is {\footnotesize\tt BW}, $C$~turns to white and $D$~remains
black. Accordingly, the locomotive created at the neighbour of~$D$ goes on its way, leaving the
round-about while that which was created close to~$C$ and~$E$ is killed.

\vtop{
\vspace{-5pt}
\ligne{\hfill\hskip 15pt
\includegraphics[scale=0.5]{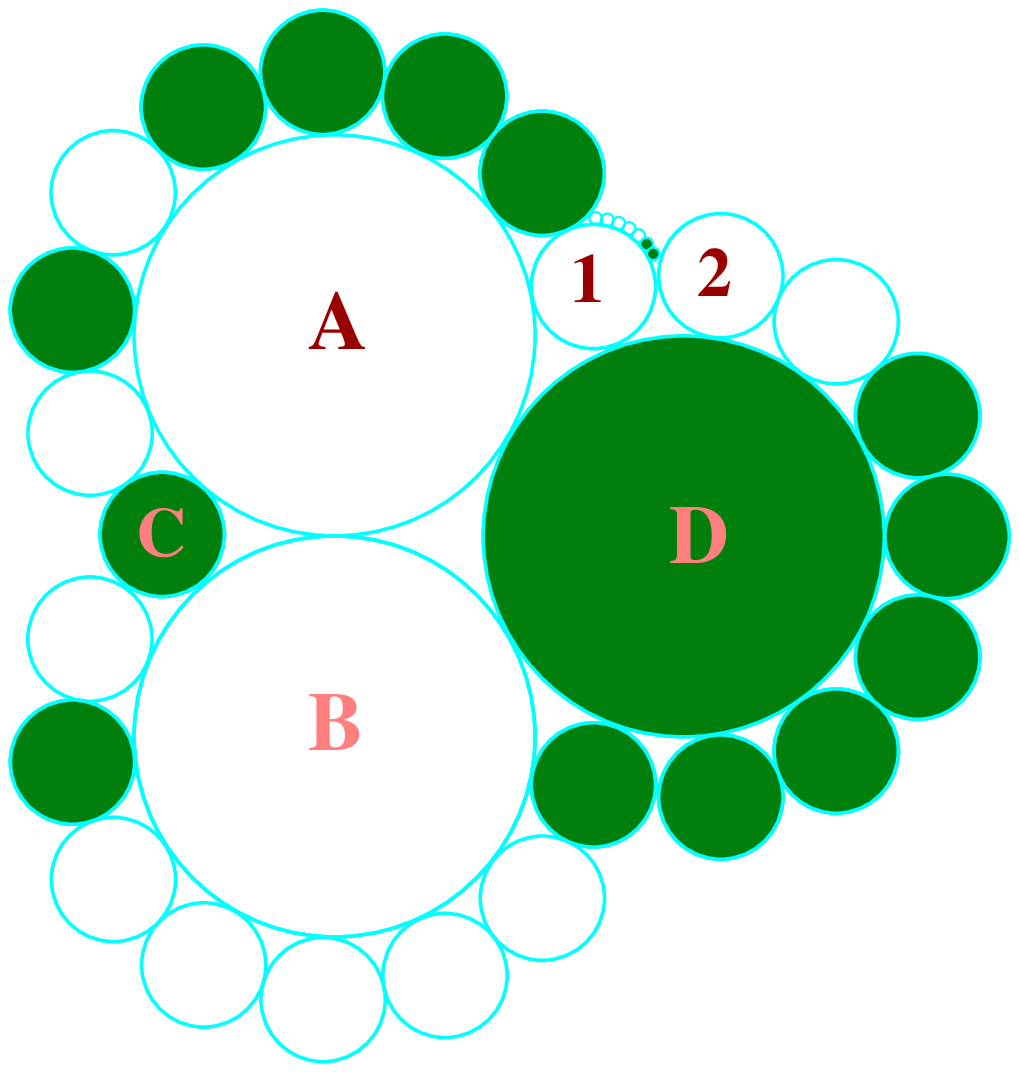}
\includegraphics[scale=0.5]{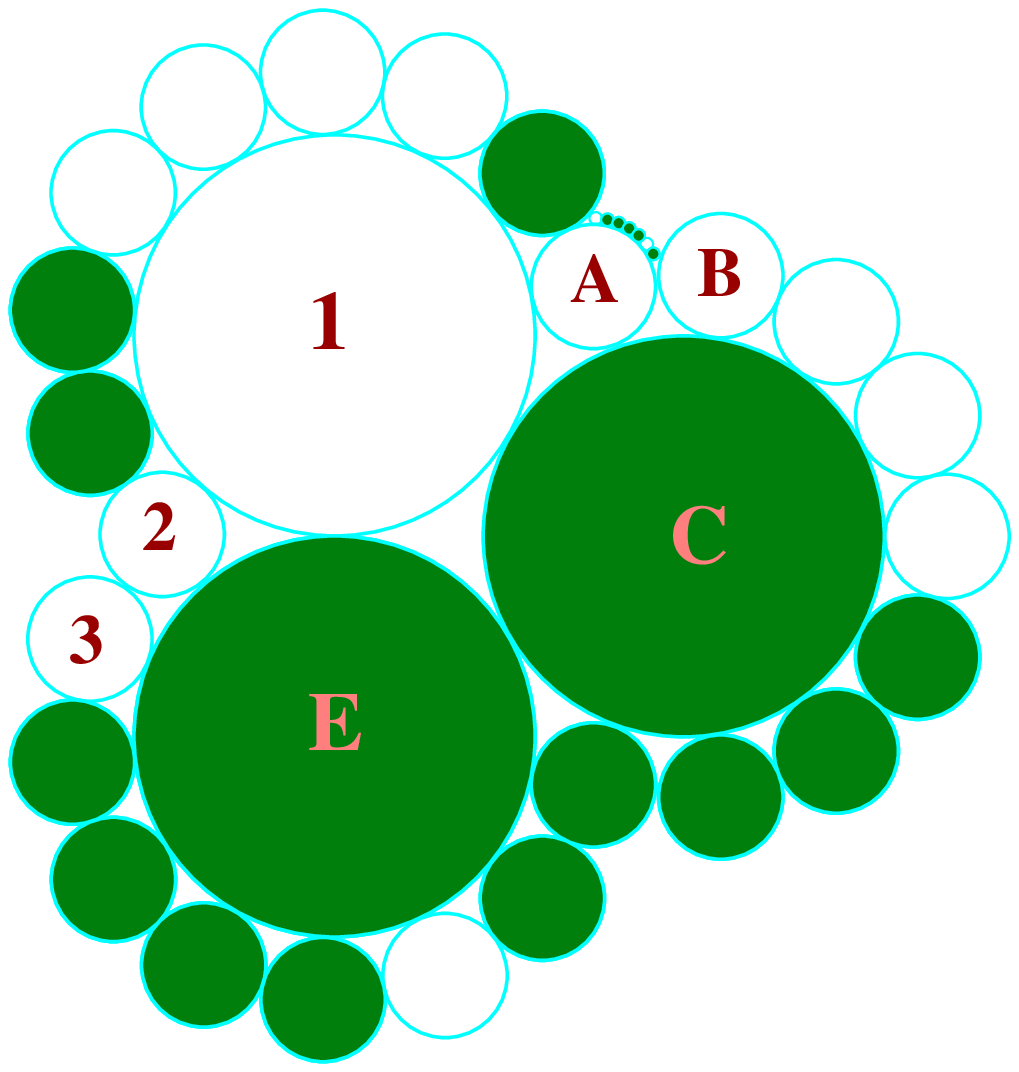}
\hfill}
\vspace{-10pt}
\begin{fig}\label{select11_zooms}
\leurre
Zooming at the idle configuration of the selector.
\end{fig}
}

Figure~\ref{select11_zooms} zooms at~$D$ and~$C$ which allows us to see more clearly how
the crossing is processed. Note that $E$ plays a role: the locomotive created at~1, right-hand side
of the figure arrives at~2 when $C$~becomes white. So that $E$ has to become white when~$C$
is flashing. This allows us to kill the locomotive sent in this direction.

\subsubsection*{The active switches}

From the structure of the selector, we can easily derive a structure which
we call the {\bf fork} which receives one locomotive and dispatches a copy of it two
different directions. 

For the convenience of the reader, we reproduce the picture in
the left-hand side part of Figure~\ref{switchactive}.
We remind the reader that this structure is used in the flip-flop, the active memory
switch and in the passive memory switch too, see Figures~\ref{flipflop},
\ref{memoactive} and~\ref{memopassive}. In the right-hand side part of the same figure,
we illustrate the controlling device used by both the flip-flop
and the active memory switch, again look at Figures~\ref{flipflop} and~\ref{memoactive}.

   In the right-hand side part of Figure~\ref{switchactive}, we zoom at the cell~{\bf c}
which looks at the passage of the locomotive. If the cell is white, the locomotive is allowed
to pass and, necessarily, it passes: in this case, the cell~{\bf t} which is on the track followed
by the locomotive has the neighbourhood of an element of the track. If the cell is black, then
it prevents the locomotive from entering the cell~{\bf t} so that the locomotive is killed.

\vtop{
\vspace{-10pt}
\ligne{\hfill\hskip 15pt
\includegraphics[scale=0.45]{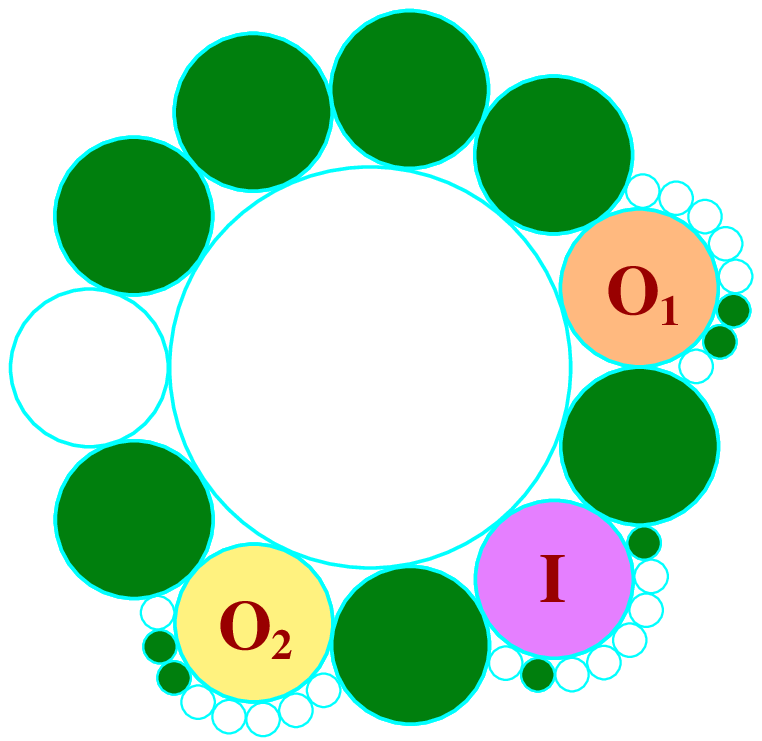}
\hskip 30pt\raise-20pt\hbox{\includegraphics[scale=0.5]{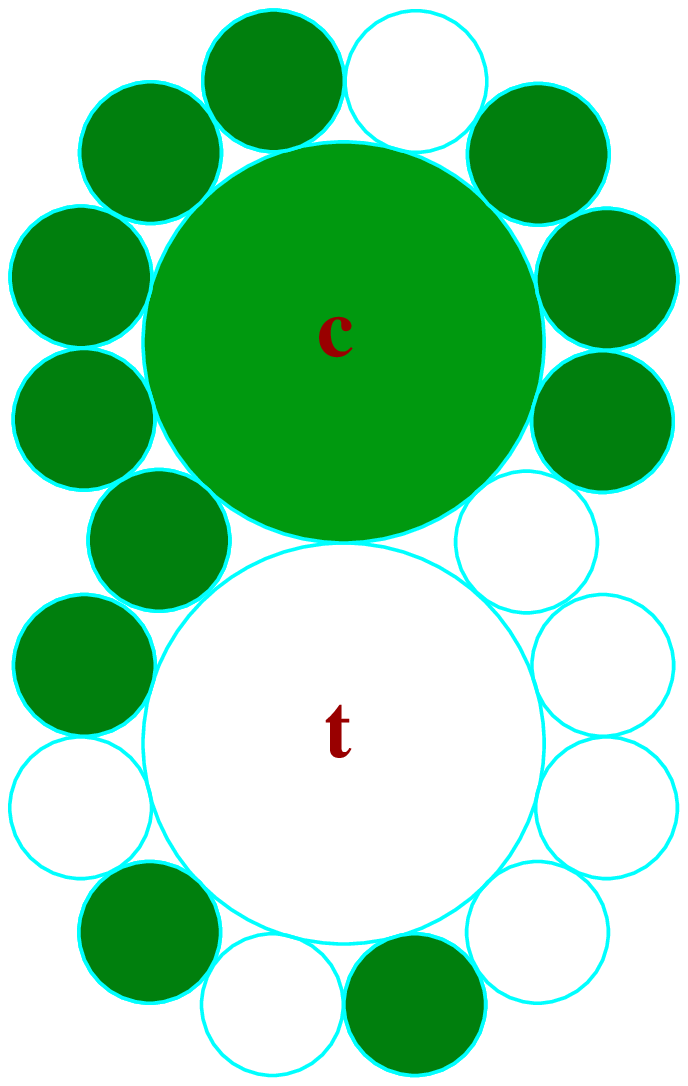}}
\hfill}
\vspace{-10pt}
\begin{fig}\label{switchactive}
\leurre
The idle configuration of the fork and of the controller used in the active switches. 
\end{fig}
}

\subsubsection*{The passive memory switch}

From Subsection~\ref{genpatt}, we know that the passive memory switch requires a more
complex structure than the passive one: here we have rather a {\bf sensor} than a controller.
The reason is that the sensor does not stop the locomotive.

\vtop{
\vspace{-20pt}
\ligne{\hfill\hskip 15pt
\includegraphics[scale=0.45]{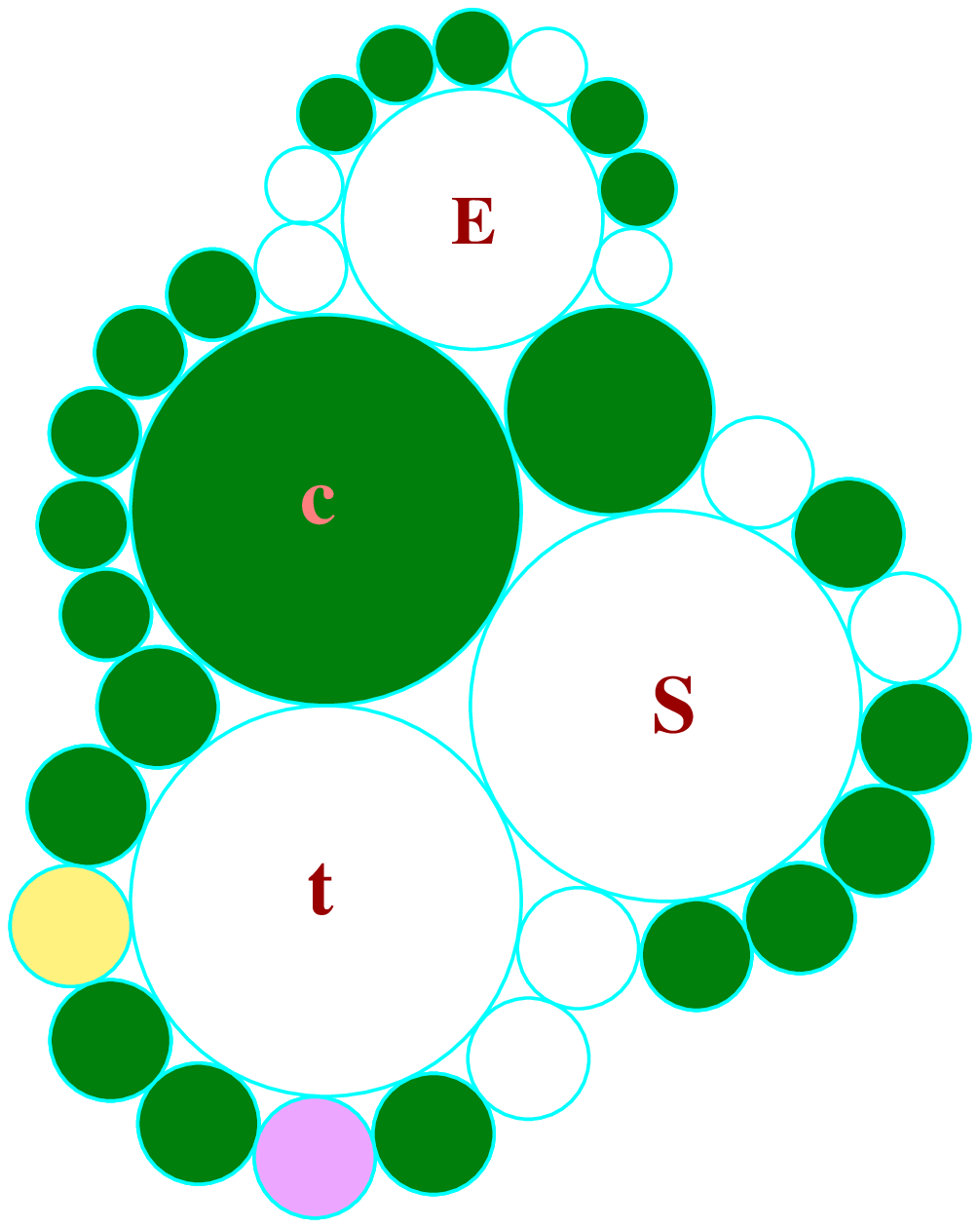}
\hfill}
\begin{fig}\label{sensor11}
\leurre
The sensor used by the passive memory switch.
\end{fig}
}

However, the sensor is
a more active structure: if it is black and if a locomotive passes through the cell~{\bf t},
the cell~{\bf S} of the sensor which can see both~{\bf c} and~{\bf t} realizes that 
the locomotive is passively running on the non-selected track. So that the sensor changes the
selection: {\bf c} becomes white. But on the other sensor of the switch, the cell~{\bf c} is
also white so that second cell~{\bf c} must become black. This is why the cell~{\bf S}
sends a locomotive which reaches both the second cell~{\bf c} and also the fork of the
active switch in order to change the states in both its controllers.

   Now, the signal sent by one of the sensors to the other enters the cell~{\bf E} which
turns the white state of the cell~{\bf c} to black. 

\section{Checking the automaton: the rules}
\label{rules}
\def\uu #1{{\footnotesize\tt #1}}
\definecolor{red}{rgb}{1.00,0.00,0.00}

   In this section, we give the rules used by the automaton and we prove them.
We first describe the format used for the rules and then we shall provide the rules used for
each configuration. We illustrate the rules by figures describing the motion
of the locomotive in the different situations described in Section~\ref{scenar} and, especially,
in Subsection~\ref{implements}. The rules are numbered, which will allow us to follow there
application in the scheduling tables of the section. These tables provides us with
the state of the key cells in the circuit at different times together with the rule which was 
applied at that time for this cell.

   First, we fix the format of the rules. To this purpose, we number the sides of the cell and we
say that the side~{\bf i} is shared by the cell and by its neighbour~{\bf i}. We shall consider
that {\bf i}$~\in~\{1..11\}$ and that the numbers increase while counter-clockwise turning around
a cell. As here we construct a rotation invariant automaton, it is not important to fix which
side is side~{\bf 1}. For instance, considering Figure~\ref{elem_voie}, the rule applied to
the cell of the track for the leftmost cell in the upper row will be denoted
\hbox{\footnotesize\tt$\underline{\hbox{\tt W}}$BBWBWBWWWWW$\underline{\hbox{\tt W}}$}. In this 
format, in the non underlined part of the word, the $i^{\rm th}$ letter from the left indicates the
state of the neighbour~{\bf i}: this can easily be checked on Figure~\ref{elem_voie}. This
format will be that of the rules which will be displayed from now on. In the word denoting a
rule, the non-underlined part is called the {\bf context} of the rule: it is the list of the
sates of the neighbours, from~1 to~11.

   Second: before listing the rules we have to note that the role of many rules consists in
keeping a considered configuration persistent. By this we mean that after a possible modification
introduced by the passage of the locomotive, the configuration recovers a state it keeps most of
the time or, at least, for a long time. As an example of the second situation we have the 
controllers of the flip-flop: the indication of whether the passage by the locomotive is allowed
or not depends on the selected track which may be changed but, once the selection is fixed, the
indication remains permanent until a possibly new one is fixed, much later if we consider the
number of steps of the automaton. In many cases, the configurations are marked by black cells,
{\tt B} in the rules while the background of the space is white, marked by {\tt W} in the rules.
In most situations, the white cells of the background have at most two contiguous neighbours.
So that the first ten rules of the automaton are:

\newcount\regle\regle=1
\def\iii #1#2#3 {\ligne{\hfill\footnotesize\tt\the\regle\hskip 5pt$\underline{\hbox{\tt#1}}$%
#2$\underline{\hbox{\tt#3}}$\hfill}\vskip-4pt
\global\advance\regle by 1}
\vtop{
\begin{tab}\label{conserv}
\leurre
The first rules: conservative rules for the milestones.
\end{tab}
\vspace{-12pt}
\ligne{\hfill
\vtop{\hsize=76pt
\iii {W}{WWWWWWWWWWW}{W}
\iii {B}{WWWWWWWWWWW}{B}
\iii {W}{BWWWWWWWWWW}{W}
\iii {B}{BWWWWWWWWWW}{B}
}
\hskip 15pt
\vtop{\hsize=76pt
\iii {W}{BBWWWWWWWWW}{W}
\iii {B}{BBWWWWWWWWW}{B}
\iii {W}{BWBWWWWWWWW}{W}
\iii {B}{BWBWWWWWWWW}{B}
}
\hskip 15pt
\vtop{\hsize=76pt
\iii {W}{BBBWWWWWWWW}{W}
\iii {B}{BBBWWWWWWWW}{B}
}
\hfill}
}

\subsection{Motion rules}

   As Figure~\ref{elem_voie} represents the idle configuration only, we here provide the reader with
an illustration of the motion of one or two locomotives through an element of the tracks,
see Figures~\ref{elem_voie_movs} and~\ref{elem_voie_movd}. This will allow the reader to 
easier follow the
checking of the rules given by Table~\ref{motion}.
It is not superfluous to remind the reader that several rules of Table~\ref{conserv}
are also applied in this situation.

\vtop{
\vspace{-10pt}
\ligne{\hfill
\includegraphics[scale=0.3]{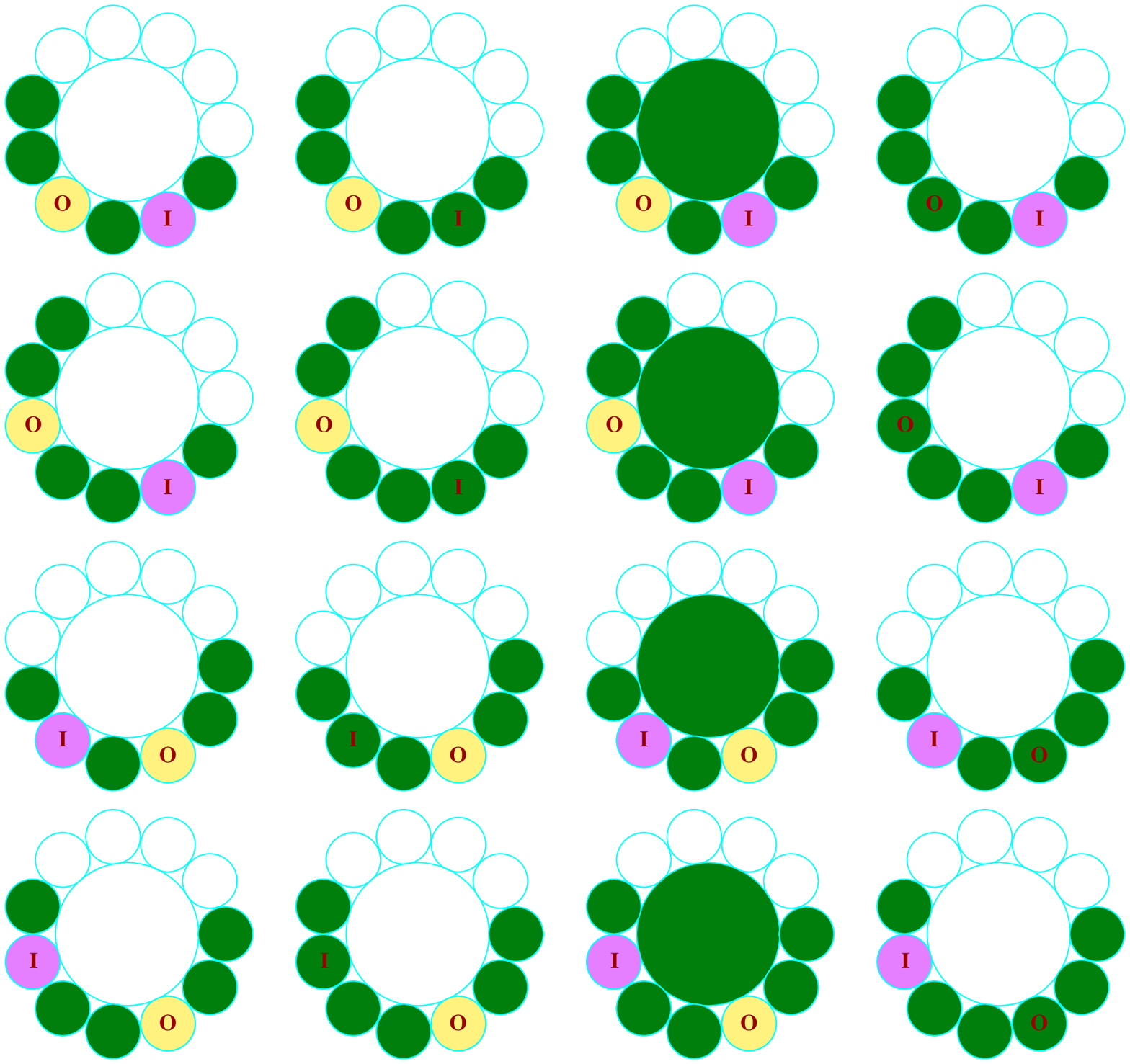}
\hfill}
\vspace{-10pt}
\begin{fig}\label{elem_voie_movs}
\leurre
The four possible motions for the elements of the tracks. Here, for a single locomotive.
\end{fig}
}

   Let us look carefully at the situation. With Figure~\ref{elem_voie} in mind, fix a cell
of the track.
Let us consider that side~{\bf 1} is the leftmost black neighbour of the cell. Then, the 
rules of Table~\ref{motion} apply to the cell of the track. Rules~11 and~18, 28 and~34
apply to an
{\bf idle configuration}: this means that the locomotive is far away from that cell, so that
if it is white, it remains white for the next time. For the milestones of the track, we can see
that in an idle configuration rules~4 applies. For neighbours of a milestone, 
rules~3 and~5 apply
while rule~9 applies to the neighbours of the cell which are its entrance and its exit for the
locomotive. Rule~10 applies to some milestones when the locomotive is in the cell.

\vtop{
\vspace{-10pt}
\ligne{\hfill
\includegraphics[scale=0.3]{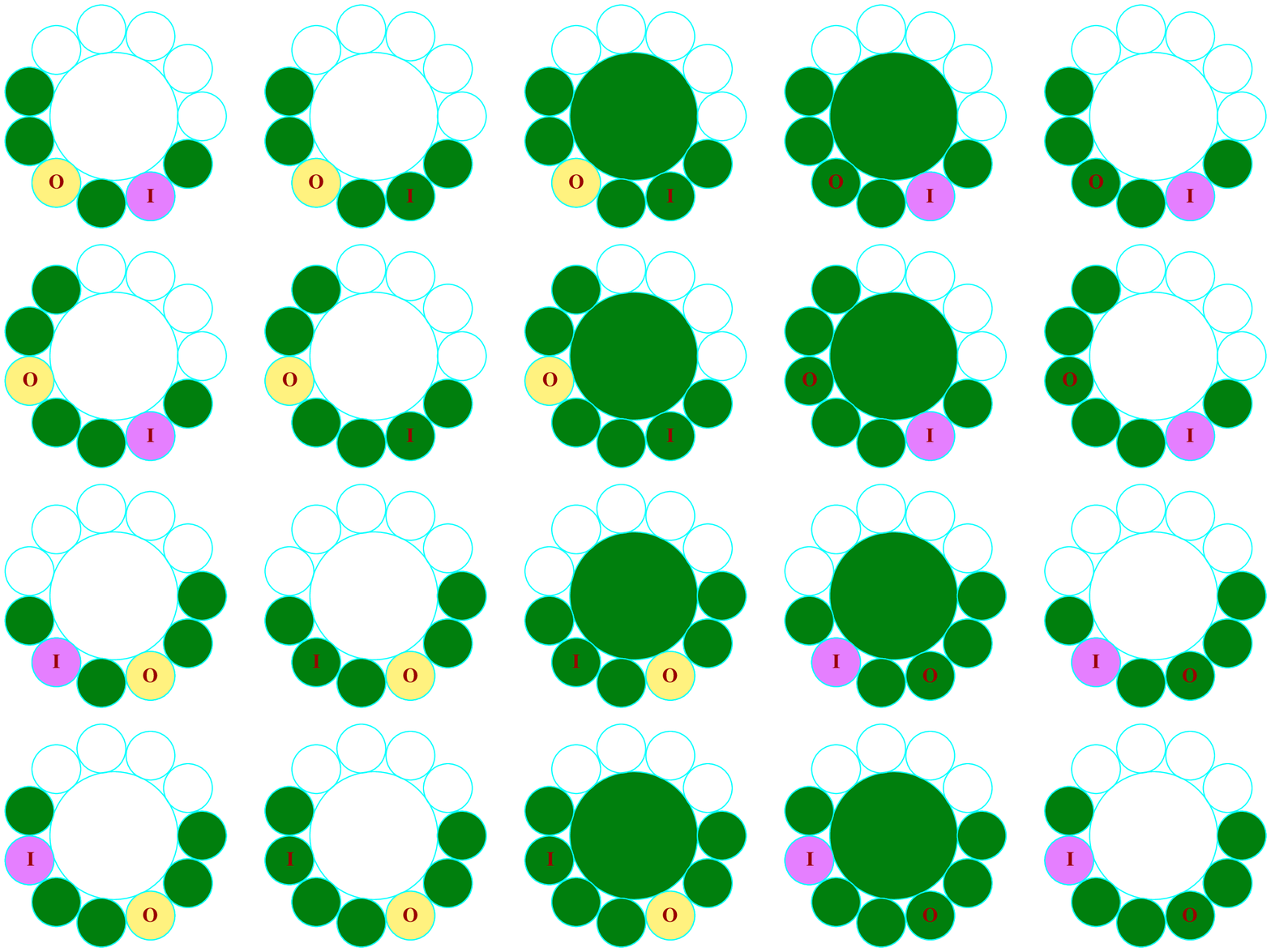}
\hfill}
\vspace{-10pt}
\begin{fig}\label{elem_voie_movd}
\leurre
The four possible motions for the elements of the tracks
for two contiguous locomotives.
\end{fig}
}

   First, consider a cell with a single pivot. 
When a single locomotive crosses it, the rules 
which apply to the cell
are rules~11, 12, 13 and~14%
{} in one direction, and rules 23, 24, 25 and~26
{} in the opposite direction.
When two contiguous locomotives cross the cell, rules~12 or~24 apply as the cell see the first
locomotive. When rules~12 and~24 are applied, 
the first 
locomotive occupies the cell while
the second one is a neighbour of the cell, 
the neighbour which was occupied by the first locomotive
at the previous time. This is why rules~15 and~27 
apply to the first locomotive. At the next time,
the second locomotive occupies the cell while the first one already left it and occupies a 
neighbour of the cell, neighbour~3{}  or~4, depending on the direction of the motion.
Next, rules~16 and~28 apply to the second locomotive, making it to leave the cell. The cell can see
the first locomotive in neighbour~3 or~4, depending on which type of cell,
and it returns to white. Accordingly, at the next step, the cell is white and the second locomotive
is in neighbour~3 or~4. As the cell has to remain white, rules~16 and~28 apply. At the next time,
the configuration is idle again so that rules~11 and~23 apply to the cell.

Note that a whole train of consecutive 
locomotives could cross the cell using the rules for the case of two consecutive locomotives.

   For a cell in between two pivots, the principle is exactly the same. The difference is that
instead of one black neighbour between the entrance and the exit, there are two consecutive black
neighbours.

\vtop{
\vspace{-10pt}
\begin{tab}\label{motion}
\leurre
The motion rules: for a single locomotive and when two contiguous locomotives travel
on the tracks. In each case, motion in one direction and motion in the opposite one.
\end{tab}
\vspace{-12pt}
\ligne{\hfill
\vtop{\hsize 152pt
\footnotesize 
\ligne{\hfill from left to right\hfill}
}
\hfill
\vtop{\hsize 152pt
\footnotesize 
\ligne{\hfill from right to left\hfill}
}
\hfill}
\vskip 0pt
\ligne{\hfill
\vtop{\hsize 76pt
\footnotesize 
\ligne{\hfill one pivot\hfill}
}
\hfill
\vtop{\hsize 76pt
\footnotesize 
\ligne{\hfill in between\hfill}
}
\vtop{\hsize 76pt
\footnotesize 
\ligne{\hfill one pivot\hfill}
}
\hfill
\vtop{\hsize 76pt
\footnotesize 
\ligne{\hfill in between\hfill}
}
\hfill}
\vskip 0pt
\ligne{\hfill
\vtop{\hsize 76pt
\ligne{\footnotesize\hskip 24pt i\hskip 6pt o\hfill}
\vskip -3pt
\iii {W}{BWBWBBWWWWW}{W}
\iii {W}{BBBWBBWWWWW}{B}
\iii {B}{BWBWBBWWWWW}{W}
\iii {W}{BWBBBBWWWWW}{W}
\vskip 2pt
\ligne{\hfill\uu {2 locomotives}\hfill}
\vskip-1pt
\iii {B}{BBBWBBWWWWW}{B}
\iii {B}{BWBBBBWWWWW}{W}
}
\hfill
\vtop{\hsize 76pt
\ligne{\footnotesize\hskip 25pt i\hskip 10pt o\hfill}
\vskip -3pt
\iii {W}{BWBBWBBWWWW}{W}
\iii {W}{BBBBWBBWWWW}{B}
\iii {B}{BWBBWBBWWWW}{W}
\iii {W}{BWBBBBBWWWW}{W}
\vskip 2pt
\ligne{\hfill\uu {2 locomotives}\hfill}
\vskip-1pt
\iii {B}{BBBBWBBWWWW}{B}
\iii {B}{BWBBBBBWWWW}{W}
}
\hfill
\vtop{\hsize 76pt
\ligne{\footnotesize\hskip 29pt o\hskip 5pt i\hfill}
\vskip -3pt
\iii {W}{BBWBWBWWWWW}{W}
\iii {W}{BBWBBBWWWWW}{B}
\iii {B}{BBWBWBWWWWW}{W}
\iii {W}{BBBBWBWWWWW}{W}
\vskip 2pt
\ligne{\hfill\uu {2 locomotives}\hfill}
\vskip-1pt
\iii {B}{BBWBBBWWWWW}{B}
\iii {B}{BBBBWBWWWWW}{W}
}
\hfill
\vtop{\hsize 76pt
\ligne{\footnotesize\hskip 28pt o\hskip 10pt i\hfill}
\vskip -3pt
\iii {W}{BBWBBWBWWWW}{W}
\iii {W}{BBWBBBBWWWW}{B}
\iii {B}{BBWBBWBWWWW}{W}
\iii {W}{BBBBBWBWWWW}{W}
\vskip 2pt
\ligne{\hfill\uu {2 locomotives}\hfill}
\vskip-1pt
\iii {B}{BBWBBBBWWWW}{B}
\iii {B}{BBBBBWBWWWW}{W}
}
\hfill
}
}
\vskip 15pt

\newdimen\decal\decal=5pt
\def\ennoir #1{%
\hbox{\hskip \decal\color{red}{\tt\bf #1}}
}

\newdimen\taille\taille=18pt
\def\prerangee #1 #2 #3 #4 #5 #6 {%
\hbox to \taille{\hfill #1\hfill}
\hbox to \taille{\hfill #2\hfill}
\hbox to \taille{\hfill #3\hfill}
\hbox to \taille{\hfill #4\hfill}
\hbox to \taille{\hfill #5\hfill}
\hbox to \taille{\hfill #6\hfill}
}
\def\prerangeev #1 #2 #3 #4 #5 {%
\hbox to \taille{\hfill #1\hfill}
\hbox to \taille{\hfill #2\hfill}
\hbox to \taille{\hfill #3\hfill}
\hbox to \taille{\hfill #4\hfill}
\hbox to \taille{\hfill #5\hfill}
}
\def\prerangeeiv #1 #2 #3 #4 {%
\hbox to \taille{\hfill #1\hfill}
\hbox to \taille{\hfill #2\hfill}
\hbox to \taille{\hfill #3\hfill}
\hbox to \taille{\hfill #4\hfill}
}
\def\prerangeeiii #1 #2 #3 {%
\hbox to \taille{\hfill #1\hfill}
\hbox to \taille{\hfill #2\hfill}
\hbox to \taille{\hfill #3\hfill}
}
\def\prerangeeii #1 #2 {%
\hbox to \taille{\hfill #1\hfill}
\hbox to \taille{\hfill #2\hfill}
}
\def\prerangeei #1 {%
\hbox to \taille{\hfill #1\hfill}
}

\vtop{\taille=13pt\decal=1pt
\vspace{-10pt}
\begin{tab}\label{schedule_voies}
\leurre
Motion of a single locomotive. To left: motion from left to right. To right: motion from right to
left. 
\end{tab}
\vskip-9pt
\ligne{\hskip 15pt
\vtop{\footnotesize\tt\hsize=200pt
\ligne{\prerangee {} 1 2 3 4 5 \prerangeeiii 6 7 8 
\hfill}
\ligne{\prerangee {1} {\ennoir {13}} {12} {11} {17} {17}  \prerangeeiii {11} {11} {14} \hfill}
\vskip-2pt
\ligne{\prerangee {2} {14} {\ennoir {13}} {12} {17} {17}  \prerangeeiii {11} {11} {11} \hfill}
\vskip-2pt
\ligne{\prerangee {3} {11} {14} {\ennoir {13}} {18} {17}  \prerangeeiii {11} {11} {11} \hfill}
\vskip-2pt
\ligne{\prerangee {4} {11} {11} {14} {\ennoir {19}} {18}  \prerangeeiii {11} {11} {11} \hfill}
\vskip-2pt
\ligne{\prerangee {5} {11} {11} {11} {20} {\ennoir {19}}  \prerangeeiii {12} {11} {11} \hfill}
\vskip-2pt
\ligne{\prerangee {6} {11} {11} {11} {17} {20}  \prerangeeiii {\ennoir {13}} {12} {11} \hfill}
\vskip-2pt
\ligne{\prerangee {7} {11} {11} {11} {17} {17}  \prerangeeiii {14} {\ennoir {13}} {12} \hfill}
\vskip-2pt
\ligne{\prerangee {8} {12} {11} {11} {17} {17}  \prerangeeiii {11} {14} {\ennoir {13}} \hfill}
}
\hskip-30pt
\vtop{\footnotesize\tt\hsize=200pt
\ligne{\prerangee {} 1 2 3 4 5 \prerangeeiii 6 7 8 
\hfill}
\ligne{\prerangee {1} {26} {23} {23} {29} {29}  \prerangeeiii {23} {24} {\ennoir {25}} \hfill}
\vskip-2pt
\ligne{\prerangee {2} {23} {23} {23} {29} {29}  \prerangeeiii {24} {\ennoir {25}} {26} \hfill}
\vskip-2pt
\ligne{\prerangee {3} {23} {23} {23} {29} {30}  \prerangeeiii {\ennoir {25}} {26} {23} \hfill}
\vskip-2pt
\ligne{\prerangee {4} {23} {23} {23} {30} {\ennoir {31}}  \prerangeeiii {26} {23} {23} \hfill}
\vskip-2pt
\ligne{\prerangee {5} {23} {23} {24} {\ennoir {31}} {32}  \prerangeeiii {23} {23} {23} \hfill}
\vskip-2pt
\ligne{\prerangee {6} {23} {24} {\ennoir {25}} {32} {29}  \prerangeeiii {23} {23} {23} \hfill}
\vskip-2pt
\ligne{\prerangee {7} {24} {\ennoir {25}} {26} {29} {29}  \prerangeeiii {23} {23} {23} \hfill}
\vskip-2pt
\ligne{\prerangee {8} {\ennoir {25}} {26} {23} {29} {29}  \prerangeeiii {23} {23} {24} \hfill}
}
\hfill}
}

\vtop{\taille=13pt\decal=1pt
\begin{tab}\label{schedule_voied}
\leurre
Motion of two contiguous locomotives. To left: motion from left to right. To right: motion 
from right to left. As shown in the last line, cells~$1$ and~$8$ are neighbours on the tracks too.
\end{tab}
\vskip-9pt
\ligne{\hskip 15pt
\vtop{\footnotesize\tt\hsize=200pt
\ligne{\prerangee {} 1 2 3 4 5 \prerangeeiii 6 7 8 
\hfill}
\ligne{\prerangee {1} {\ennoir {16}} {\ennoir {15}} {12} {17} {17}  \prerangeeiii {11} {11} {14} \hfill}
\vskip-2pt
\ligne{\prerangee {2} {14} {\ennoir {16}} {\ennoir {15}} {18} {17}  \prerangeeiii {11} {11} {11} \hfill}
\vskip-2pt
\ligne{\prerangee {3} {11} {14} {\ennoir {16}} {\ennoir {21}} {18}  \prerangeeiii {11} {11} {11} \hfill}
\vskip-2pt
\ligne{\prerangee {4} {11} {11} {14} {\ennoir {22}} {\ennoir {21}}  \prerangeeiii {12} {11} {11} \hfill}
\vskip-2pt
\ligne{\prerangee {5} {11} {11} {11} {20} {\ennoir {22}}  \prerangeeiii {\ennoir {15}} {12} {11} \hfill}
\vskip-2pt
\ligne{\prerangee {6} {11} {11} {11} {17} {20}  \prerangeeiii {\ennoir {16}} {\ennoir {15}} {12} \hfill}
\vskip-2pt
\ligne{\prerangee {7} {12} {11} {11} {17} {17}  \prerangeeiii {14} {\ennoir {16}} {\ennoir {15}} \hfill}
\vskip-2pt
\ligne{\prerangee {8} {\ennoir {15}} {12} {11} {17} {17}  \prerangeeiii {11} {14} {\ennoir {16}} \hfill}
}
\hskip-30pt
\vtop{\footnotesize\tt\hsize=200pt
\ligne{\prerangee {} 1 2 3 4 5 \prerangeeiii 6 7 8 
\hfill}
\ligne{\prerangee {1} {26} {23} {23} {29} {29}  \prerangeeiii {24} {\ennoir {27}} {\ennoir {28}} \hfill}
\vskip-2pt
\ligne{\prerangee {2} {23} {23} {23} {29} {30}  \prerangeeiii {\ennoir {27}} {\ennoir {28}} {26} \hfill}
\vskip-2pt
\ligne{\prerangee {3} {23} {23} {23} {30} {\ennoir {33}}  \prerangeeiii {\ennoir {28}} {26} {23} \hfill}
\vskip-2pt
\ligne{\prerangee {4} {23} {23} {24} {\ennoir {33}} {\ennoir {34}}  \prerangeeiii {26} {23} {23} \hfill}
\vskip-2pt
\ligne{\prerangee {5} {23} {24} {\ennoir {27}} {\ennoir {34}} {32}  \prerangeeiii {23} {23} {23} \hfill}
\vskip-2pt
\ligne{\prerangee {6} {24} {\ennoir {27}} {\ennoir {28}} {32} {29}  \prerangeeiii {23} {23} {23} \hfill}
\vskip-2pt
\ligne{\prerangee {7} {\ennoir {27}} {\ennoir {28}} {26} {29} {29}  \prerangeeiii {23} {23} {24} \hfill}
\vskip-2pt
\ligne{\prerangee {8} {\ennoir {28}} {26} {23} {29} {29}  \prerangeeiii {23} {24} {\ennoir {27}} \hfill}
}
\hfill}
}
\vskip 10pt
   Tables~\ref{schedule_voies} and~\ref{schedule_voied} illustrate the motion in the following
way. Eight consecutive cells of the track are taken in one direction in 
Table~\ref{schedule_voies}, in the other direction in Table~\ref{schedule_voied}. In both
cases, cells are numbered from~1 to~8 and cells~4 and~5 are in between two consecutive pivots.
The tables indicate for each cell and for each time which instruction applies. When the number
of the instruction is in \ennoir{red bold digits}, this means that the corresponding cell is black
before the rule is applied, otherwise it is white.
Figure~\ref{voie11_1} allows us to check this. Note that Tables~\ref{schedule_voies} 
and~\ref{schedule_voied} were filled up by a computer program simulating the automaton.
Later on, such tables will be called {\bf schedule tables}.

%
%
\subsection{Fixed switch}

    The fixed switch has been described in Sub-section~\ref{implements}.

   Figure~\ref{fix11_mov} illustrates the motion of one and two contiguous locomotives
through a fixed switch. The rules are given in Table~\ref{fix_rules}. Basically, the motion is that
of a locomotive through an element of the track. The first two lines illustrate the passive passage
of a single locomotive. In the first row, the locomotive comes from the left, in the second row, it
comes from the right. The rule for keeping the idle configuration is rule~35.
Rule~36 allows the locomotive to enter the locomotive from the left. Rule~39 does the same
for a locomotive coming from the right.

\vtop{
\vspace{-5pt}
\ligne{\hfill\hskip 15pt
\includegraphics[scale=0.4]{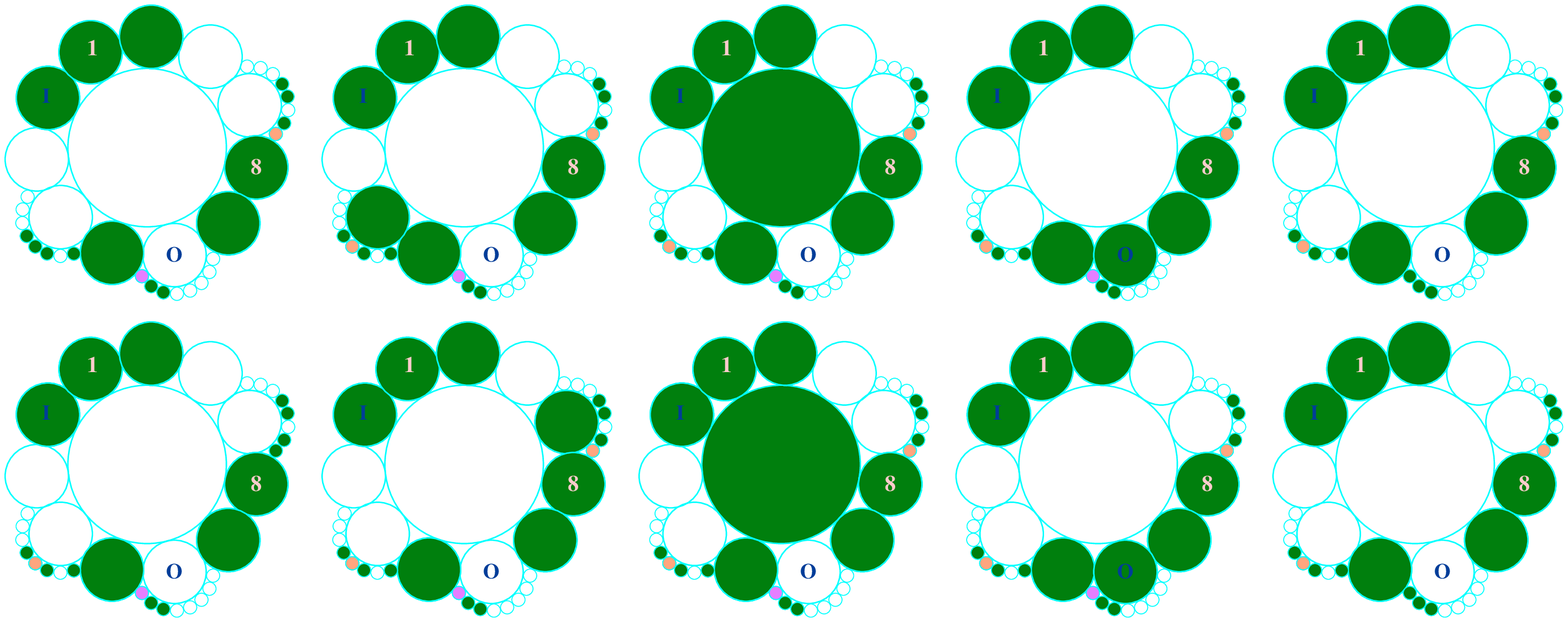}
\hfill}
\vspace{-5pt}
\ligne{\hfill\hskip 15pt
\includegraphics[scale=0.4]{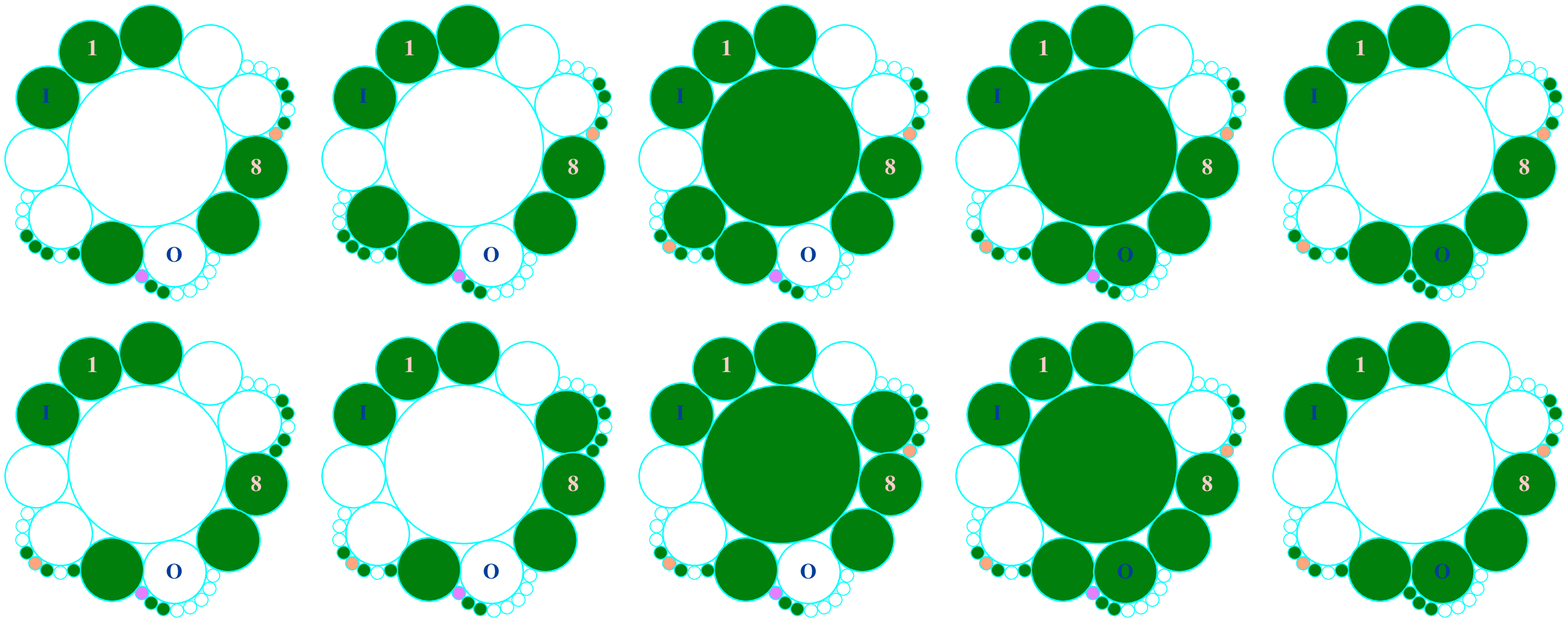}
\hfill}
\begin{fig}\label{fix11_mov}
\leurre
The motion of the locomotive through a fixed switch. First two rows: a single locomotive.
Last two rows: two contiguous locomotives.
\end{fig}
}
\vskip-5pt
\vtop{
\begin{tab}\label{fix_rules}
\leurre
The rules which handle a fixed switch.
\end{tab}
\vspace{-12pt}
\ligne{\hfill\footnotesize
\vtop{\hsize 76pt
\ligne{\hfill a single locomotive\hfill}
}
\hskip 15pt
\vtop{\hsize 76pt
\ligne{\hfill two locomotives\hfill}
}
\hskip 15pt
\vtop{\hsize 76pt
\ligne{\hfill tracks again\hfill}
}
\hfill}
\vskip 0pt
\ligne{\hfill
\footnotesize\tt
\vtop{\hsize 76pt
\ligne{\footnotesize\hskip 38pt i$_1$\hskip 2pt o\hskip 8pt i$_2$\hfill}
\vskip-3pt
\iii {W}{BBBWWBWBBWW}{W}
\iii {W}{BBBWBBWBBWW}{B}
\iii {B}{BBBWWBWBBWW}{W}
\iii {W}{BBBWWBBBBWW}{W}
\vskip 3pt
\iii {W}{BBBWWBWBBBW}{B}
}
\hskip 15pt
\vtop{\hsize 76pt
\ligne{\footnotesize\hskip 38pt i$_\ell$\hskip 2pt o\hskip 8pt i$_r$\hfill}
\vskip-3pt
\iii {B}{BBBWBBWBBWW}{B}
\iii {B}{BBBWWBBBBWW}{W}
\iii {B}{BBBWWBWBBBW}{B}
}
\hskip 15pt
\vtop{\hsize 76pt
\iii {W}{BWBWBBBWWWW}{W}
\iii {B}{BWBWBBBWWWW}{W}
\vskip 3pt
\iii {B}{BWBWBBWWWWB}{W}
\iii {W}{BBWBWBBWWWW}{W}
}
\hfill}   
}
\vskip 15pt
Rules~36 and~37 make the locomotive enter and then leave the cell. Rule~44 witnesses that the 
locomotive leaves the cell. In the case of two locomotives, rules~40 and~42
allow the second locomotive
to enter the cell while the first locomotive is about to leave it. There are two rules as there
are two possible entries. Rule~41 makes the second 
locomotive leave the cell. Tables~\ref{schedule_fixes} and~\ref{schedule_fixed} are
the schedule tables illustrating the crossing of the switch by the locomotive(s). 
Here, the center of the
switch is denoted by~{\footnotesize\tt F}, the entrances by~{\footnotesize\tt i$_\ell$}
for the left-hand side one and by~{\footnotesize\tt i$_r$} for the right-hand side one.
The exit is denoted by~{\footnotesize\tt o}. For each exit/entrance cell, its neighbour of the
track is indicated as {\footnotesize\tt v$_\ell$}, {\footnotesize\tt v$_\ell$} 
and~{\footnotesize\tt w}, respectively. As the cells {\uu i$_\ell$} and {\uu i$_r$} which 
are cells of the tracks are used with a different exit from what is normally achieved, we need
two additional rules for each cell: rules~43 and~44 for the left-hand side and rules~45 and~46
for the right-hand one.

\vtop{\taille=13pt\decal=1pt
\begin{tab}\label{schedule_fixes}
\leurre
Motion of a single locomotive through the fixed switch. 
To left: the locomotive arrives from the left-hand side. To right: it arrives from the right-hand 
side.
\end{tab}
\vskip-9pt
\ligne{\hskip 15pt
\vtop{\footnotesize\tt\hsize=200pt
\ligne{\prerangee {} F o w {v$_\ell$} {i$_\ell$} \prerangeeii {v$_r$} {i$_r$} 
\hfill}
\ligne{\prerangee {1} {35} {11} {23} {\ennoir {13}} {12}  \prerangeeii {23} {11} \hfill}
\vskip-2pt
\ligne{\prerangee {2} {36} {11} {23} {14} {\ennoir {13}}  \prerangeeii {23} {11} \hfill}
\vskip-2pt
\ligne{\prerangee {3} {\ennoir {37}} {12} {23} {11} {43}  \prerangeeii {23} {46} \hfill}
\vskip-2pt
\ligne{\prerangee {4} {38} {\ennoir {13}} {24} {11} {11}  \prerangeeii {23} {11} \hfill}
\vskip-2pt
\ligne{\prerangee {5} {35} {14} {\ennoir {25}} {11} {11}  \prerangeeii {23} {11} \hfill}
\vskip-2pt
\ligne{\prerangee {6} {35} {11} {23} {11} {11}  \prerangeeii {23} {11} \hfill}
\vskip-2pt
\ligne{\prerangee {7} {35} {11} {23} {11} {11}  \prerangeeii {23} {11} \hfill}
}
\hskip-40pt
\vtop{\footnotesize\tt\hsize=200pt
\ligne{\prerangee {} F o w {v$_\ell$} {i$_\ell$} \prerangeeii {v$_r$} {i$_r$} 
\hfill}
\ligne{\prerangee {1} {35} {11} {23} {11} {11}  \prerangeeii {\ennoir {25}} {12} \hfill}
\vskip-2pt
\ligne{\prerangee {2} {39} {11} {23} {11} {11}  \prerangeeii {26} {\ennoir {13}} \hfill}
\vskip-2pt
\ligne{\prerangee {3} {\ennoir {37}} {12} {23} {11} {43}  \prerangeeii {23} {46} \hfill}
\vskip-2pt
\ligne{\prerangee {4} {38} {\ennoir {13}} {24} {11} {11}  \prerangeeii {23} {11} \hfill}
\vskip-2pt
\ligne{\prerangee {5} {35} {14} {\ennoir {25}} {11} {11}  \prerangeeii {23} {11} \hfill}
\vskip-2pt
\ligne{\prerangee {6} {35} {11} {23} {11} {11}  \prerangeeii {23} {11} \hfill}
\vskip-2pt
\ligne{\prerangee {7} {35} {11} {23} {11} {11}  \prerangeeii {23} {11} \hfill}
}
\hfill}
}

\vtop{\taille=13pt\decal=1pt
\begin{tab}\label{schedule_fixed}
\leurre
Motion of two contiguous locomotives through the fixed switch. 
To left: the locomotives arrive from the left-hand side. To right: they arrive from the right-hand 
side.
\end{tab}
\vskip-9pt
\ligne{\hskip 15pt
\vtop{\footnotesize\tt\hsize=200pt
\ligne{\prerangee {} F o w {v$_\ell$} {i$_\ell$} \prerangeeii {v$_r$} {i$_r$} 
\hfill}
\ligne{\prerangee {1} {36} {11} {23} {\ennoir {16}} {\ennoir {15}}  \prerangeeii {23} {11} \hfill}
\vskip-2pt
\ligne{\prerangee {2} {\ennoir {40}} {12} {23} {14} {\ennoir {44}}  \prerangeeii {23} {46} \hfill}
\vskip-2pt
\ligne{\prerangee {3} {\ennoir {41}} {\ennoir {15}} {24} {11} {43}  \prerangeeii {23} {46} \hfill}
\vskip-2pt
\ligne{\prerangee {4} {38} {\ennoir {16}} {\ennoir {27}} {11} {11}  \prerangeeii {23} {11} \hfill}
\vskip-2pt
\ligne{\prerangee {5} {35} {14} {\ennoir {25}} {11} {11}  \prerangeeii {23} {11} \hfill}
\vskip-2pt
\ligne{\prerangee {6} {35} {11} {23} {11} {11}  \prerangeeii {23} {11} \hfill}
\vskip-2pt
\ligne{\prerangee {7} {35} {11} {23} {11} {11}  \prerangeeii {23} {11} \hfill}
}
\hskip-40pt
\vtop{\footnotesize\tt\hsize=200pt
\ligne{\prerangee {} F o w {v$_\ell$} {i$_\ell$} \prerangeeii {v$_r$} {i$_r$} 
\hfill}
\ligne{\prerangee {1} {39} {11} {23} {11} {11}  \prerangeeii {\ennoir {28}} {\ennoir {15}} \hfill}
\vskip-2pt
\ligne{\prerangee {2} {\ennoir {42}} {12} {23} {11} {43}  \prerangeeii {26} {\ennoir {45}} \hfill}
\vskip-2pt
\ligne{\prerangee {3} {\ennoir {41}} {\ennoir {15}} {24} {11} {43}  \prerangeeii {23} {46} \hfill}
\vskip-2pt
\ligne{\prerangee {4} {38} {\ennoir {16}} {\ennoir {27}} {11} {11}  \prerangeeii {23} {11} \hfill}
\vskip-2pt
\ligne{\prerangee {5} {35} {14} {\ennoir {25}} {11} {11}  \prerangeeii {23} {11} \hfill}
\vskip-2pt
\ligne{\prerangee {6} {35} {11} {23} {11} {11}  \prerangeeii {23} {11} \hfill}
\vskip-2pt
\ligne{\prerangee {7} {35} {11} {23} {11} {11}  \prerangeeii {23} {11} \hfill}
}
\hfill}
}


\subsection{The round-about}

   In this subsection, we first look at the duplicator and then at the selector.

\subsubsection*{Duplicator}

   The study of the duplicator is illustrated by Figure~\ref{dupliq11_mov} and the rules
are given by Table~\ref{dupliq_rules}. Also, Table~\ref{schedule_dupl} allows us
to check the application of the rules and the working of the duplicator. For 
Table~\ref{schedule_dupl}, note that \uu D is the centre of the duplicator, that \uu i is
the entrance for the locomotive, that \uu i$_1$ is the neighbour of~\uu i on the tracks leading
to~\uu D, that \uu o is the exit through which the two locomotives leave~\uu D and that
they go on the tracks, first through~\uu o$_1$ and then through \uu o$_2$. 

Note that the crossing of an element 
of the tracks or of a fixed switch by a single locomotive requires four steps exactly. If 
a locomotive is about to enter such a cell at time~$t$, it just left the exit at time~$t$+4.
Now, when two consecutive locomotives cross the same elements, one more time is required.
Here, we can see a similar situation: a single locomotive enters but two contiguous ones exit,
so that one more step is needed but no more, as the pattern is reduced to a specific neighbouring
of a white cell.

\vtop{
\vspace{-5pt}
\ligne{\hfill\hskip 15pt
\includegraphics[scale=0.3]{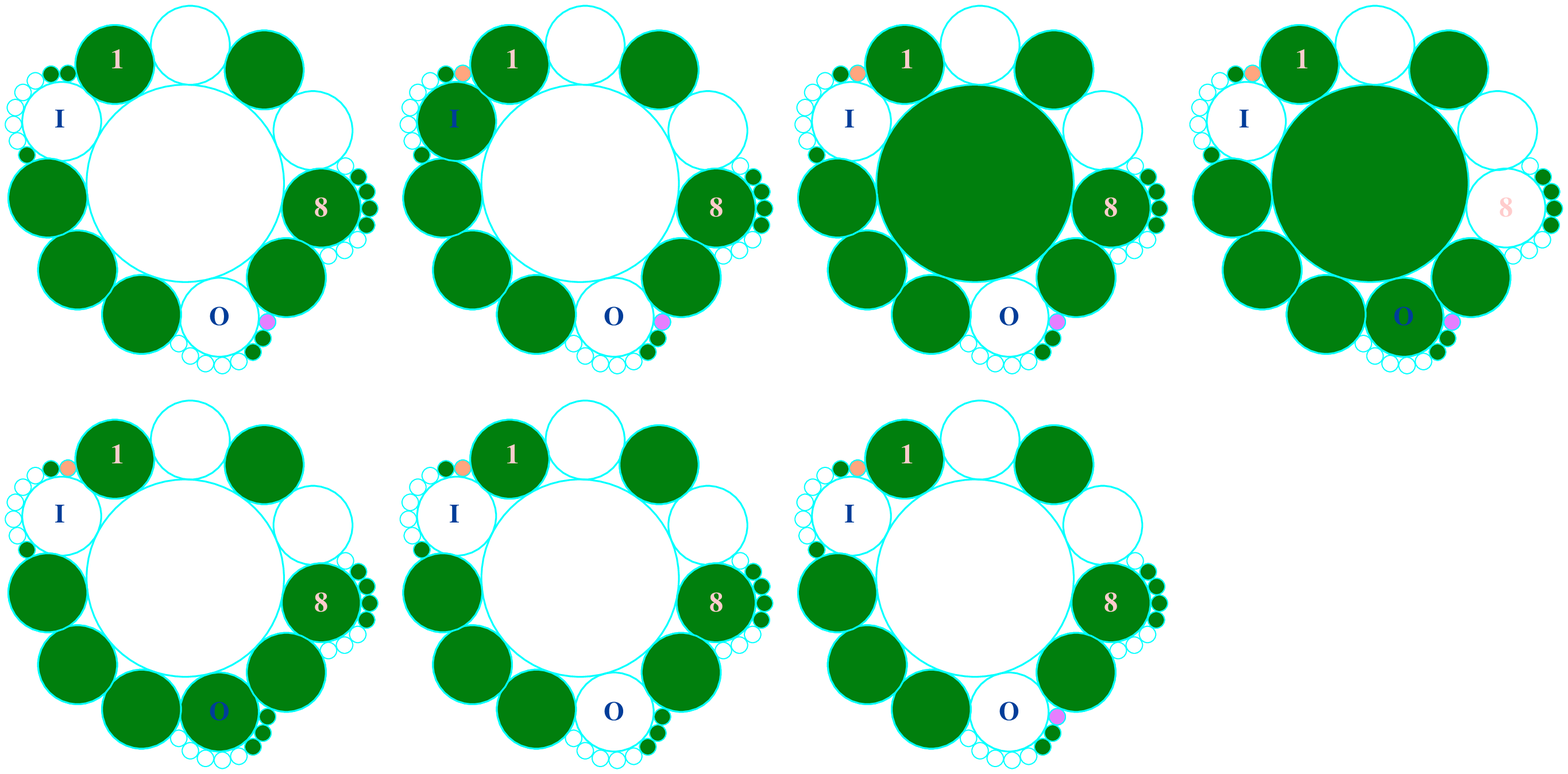}
\hfill}
\vspace{-15pt}
\begin{fig}\label{dupliq11_mov}
\leurre
The motion of the locomotive through the duplicator. One locomotive enters and two ones leave
the pattern.
\end{fig}
}

\vskip -15pt
\vtop{
\begin{tab}\label{dupliq_rules}
\leurre
The rules for the duplicator. To left: the rules for the central cell. To right, the rules
for neighbour~{\bf 8}, {\tt n$_8$}.
\end{tab}
\vspace{-12pt}
\ligne{\hfill
\footnotesize
\vtop{\hsize 76pt
\ligne{\hfill the central cell\hfill} 
}
\hskip 15pt
\vtop{\hsize 76pt
\ligne{\hfill neighbour {\bf 8}\hfill} 
}
\hfill}
\vskip-2pt
\ligne{\hfill
\footnotesize\tt
\vtop{\hsize 76pt
\ligne{\footnotesize\tt\hskip 24pt i\hskip 15pt o\hskip 4pt n$_8$\hfill}
\vskip-3pt
\iii {W}{BWBBBWBBWBW}{W}
\iii {W}{BBBBBWBBWBW}{B}
\iii {B}{BWBBBWBBWBW}{B}
\iii {B}{BWBBBBBWWBW}{W}
\iii {W}{BWBBBBBBWBW}{W}
}
\hskip 15pt
\vtop{\hsize 76pt
\ligne{\footnotesize\tt\hskip 47pt D\hfill}
\vskip-3pt
\iii {B}{BBBBWWWBWWW}{B}
\iii {B}{BBBBWWBBWWW}{W}
\iii {W}{BBBBWWBBWWW}{B}
}
\hfill} 
} 
\vskip 10pt

The creation of the second locomotive is triggered by rule~50. It introduces
a one step delay in the motion by keeping the central cell to be black. Now, rule~50 might 
produce infinitely many locomotives. In order to reduce the creation of a new locomotive to a 
single one, neighbour~{\bf 8} of the central cell flashes as the first locomotive entered~\uu D 
by rule~53 which makes it turn from black to white. Rule~54 makes neighbour~{\bf 8} return to 
the black state at the following step. When rule~53 has been applied, so that neighbour~{\bf 8} is
white, rule~50 applied to the central
cell makes the second locomotive leave the cell. Now, as shown by Tables~\ref{dupliq_rules}
and~\ref{schedule_dupl},
neighbour~{\bf 8} is ruled by rule~52. 

\vtop{\decal=3pt
\begin{tab}\label{schedule_dupl}
\leurre
Motion of the locomotive through the duplicator. {\tt D} is the central cell, {\tt i}, {\tt o}
the entrance, exit, respectively for the locomotive.
\end{tab}
\vskip-9pt
\ligne{\hskip 70pt
\vtop{\footnotesize\tt\hsize=180pt
\ligne{\prerangee {} {i$_1$} i D o {o$_1$} \prerangeeii {o$_2$} {n$_8$} 
\hfill}
\ligne{\prerangee {1} {\ennoir {25}} {24} {47} {23} {11}  \prerangeeii {11} {\ennoir {52}} \hfill}
\vskip-2pt
\ligne{\prerangee {2} {26} {\ennoir {25}} {48} {23} {11}  \prerangeeii {11} {\ennoir {52}} \hfill}
\vskip-2pt
\ligne{\prerangee {3} {23} {26} {\ennoir {49}} {24} {11}  \prerangeeii {11} {\ennoir {53}} \hfill}
\vskip-2pt
\ligne{\prerangee {4} {23} {26} {\ennoir {50}} {\ennoir {27}} {12}  \prerangeeii {11} {54} \hfill}
\vskip-2pt
\ligne{\prerangee {5} {23} {23} {51} {\ennoir {28}} {\ennoir {15}}  \prerangeeii {12} {\ennoir {52}} \hfill}
\vskip-2pt
\ligne{\prerangee {6} {23} {23} {47} {26} {\ennoir {16}}  \prerangeeii {\ennoir {15}} {\ennoir {52}} \hfill}
\vskip-2pt
\ligne{\prerangee {7} {23} {23} {47} {23} {14}  \prerangeeii {\ennoir {13}} {\ennoir {52}} \hfill}
\vskip-2pt
\ligne{\prerangee {8} {23} {23} {47} {23} {11}  \prerangeeii {11} {\ennoir {52}} \hfill}
}
\hfill}
}

\subsubsection*{Selector}

\def\xxx{\hskip 0.4pt}

   As a consequence of the task
assigned to the selector, its structure is more complex than what we have up to now studied. 
Figures~\ref{selectD_mov} and~\ref{selectE_mov} show us that
several cells around the track used by the locomotive are involved in the working of the 
structure. Tables~\ref{schedule_selects} and~\ref{schedule_selectd} show that thirteen cells are
actively involved in the working of the selector. Also, Table~\ref{select_rules} indicates
that 43 new rules are needed. Tables~\ref{schedule_selects} and~\ref{schedule_selectd} show
that 12 rules of Table~\ref{motion} are also involved. Note that globally the automaton
makes use of 115 rules, so that the selector requires almost the half of it.

The thirteen cells involved in the working of the selector are \uu i, the neighbour of~\uu B through
which the locomotive(s) enter(s) the selector after crossing \uu i$_1$, the neighbour of~\uu i
on the tracks arriving to the selector: see Figure~\ref{selectD_mov}, last line, where
the cells visited by the locomotive(s) are indicated in light colours. 
After~\uu B comes~\uu A from where the selection occurs,
controlled by~\uu C and~\uu D which both can see \uu A and~\uu B at the same time. From~\uu A,
two locomotives exit, one through~\uu o$_r$, neighboured by~\uu D, the other through~\uu o$_\ell$
which is neighboured by~\uu C. The destruction of the superfluous locomotive requires to also
examine two additional cells of the tracks on each side: \uu o$_r^1$ and \uu o$_r^2$ on the 
right hand side, \uu o$_\ell^1$ and \uu o$_\ell^2$ on the left hand side. Due to the number
of neighbours required for~\uu C, another neighbour of~\uu C, namely \uu E, is involved
in this destruction process. Note that \uu E is also a neighbour of the cells \uu o$_\ell$,
\uu o$_\ell^1$ and \uu o$_\ell^2$, while \uu C is a neighbour of \uu o$_\ell$ only.

\vskip-10pt
\vtop{
\begin{tab}\label{select_rules}
\leurre
The rules for the selector. The table gives the rules for the cells {\bf A}, {\bf D},
{\bf C} and~{\bf E}. Also see Figures~{\rm\ref{selectD_mov}} and~{\rm\ref{selectE_mov}}.
\end{tab}
\vspace{-12pt}
\ligne{\hfill
\footnotesize\tt
\vtop{\hsize 76pt
\ligne{\hfill A 
\hfill} 
}
\hfill
\vtop{\hsize 76pt
\ligne{\hfill i 
\hfill} 
}
\hfill
\vtop{\hsize 76pt
\ligne{\hfill C 
\hfill} 
}
\hfill
\hfill}
\vskip-3pt
\ligne{\hfill
\footnotesize\tt
\vtop{\hsize 76pt
\ligne{\uu{\hskip 19pt C\xxx B\xxx D\xxx o$_r$ \hskip 18pt o$_\ell$}\hfill}
\vskip-3pt
\iii {W}{BWBWBBBBWBW}{W}
\iii {W}{BBBWBBBBWBW}{B}
\iii {B}{BWBWBBBBWBW}{W}
\iii {W}{WWBBBBBBWBB}{W}
\vskip 2pt
\ligne{\hfill\uu{2 locomotives}\hfill}
\vskip-1pt
\iii {B}{BBBWBBBBWBW}{W}
\iii {W}{BWWBBBBBWBB}{W}
\vskip 2pt
\ligne{\hfill\uu{fork only:}\hfill}
\vskip-1pt
\iii {W}{BBBBBBWBBBW}{W}
}
\hfill
\vtop{\hsize 76pt
\ligne{\uu{\hskip 28pt B \hskip 10pt i$_1$}\hfill}
\vskip-3pt
\iii {W}{BBWBWBWBWWW}{W} 
\iii {W}{BBWBWBBBWWW}{B}
\iii {B}{BBWBWBWBWWW}{W}
\iii {W}{BBBBWBWBWWW}{W}
\iii {W}{BBWWWBWBWWW}{W} 
\vskip 2pt
\ligne{\hfill\uu{2 locomotives}\hfill}
\vskip-1pt
\iii {B}{BBWBWBBBWWW}{B}
\iii {B}{BBBBWBWBWWW}{W}
}
\hfill
\vtop{\hsize 76pt
\ligne{\uu{\hskip 19pt AoE \hskip 28pt B}\hfill}
\vskip-3pt
\iii {B}{WWBBBBBWWWW}{B}
\iii {B}{WWBBBBBWWBW}{B}
\iii {B}{WWBBBBBWWWB}{B}
\iii {W}{WBBBBBBWWWW}{B}
\iii {B}{WWWBBBBWWWW}{B}
\vskip 2pt
\ligne{\hfill\uu{2 locomotives}\hfill}
\vskip-1pt
\iii {B}{WWBBBBBWWBB}{B}
\iii {B}{BWBBBBBWWWB}{B}
\iii {B}{WBBBBBBWWWW}{B}
}
\hfill
}
\vskip 5pt
\ligne{\hfill
\footnotesize\tt
\vtop{\hsize 76pt
\ligne{\hfill D 
\hfill} 
}
\hfill
\vtop{\hsize 76pt
\ligne{\hfill E 
\hfill} 
}
\hfill
\vtop{\hsize 76pt
\ligne{\hfill B 
\hfill} 
}
\hfill}
\vskip-3pt
\ligne{\hfill
\footnotesize\tt
\vtop{\hsize 76pt
\ligne{\uu{\hskip 19pt B\xxx A \hskip 22pt 2\xxx 1\xxx o$_r$}\hfill}
\vskip-3pt
\iii {B}{WBBBBBBBWWW}{B}
\iii {B}{BWBBBBBBWWW}{B}
\iii {B}{WWBBBBBBWWB}{B}
\iii {B}{WWBBBBBBWBW}{B}
\vskip 2pt
\ligne{\hfill\uu{2 locomotives}\hfill}
\vskip-1pt
\iii {B}{BBBBBBBBWWW}{W}
\iii {W}{WWBBBBBBWWB}{W}
}
\hfill
\vtop{\hsize 76pt
\ligne{\uu{\hskip 19pt Co12}\hfill}
\vskip-3pt
\iii {B}{BWWWBBBBWBB}{B}
\iii {B}{WBWWBBBBWBB}{W}
\iii {W}{BWBWBBBBWBB}{B}
\vskip 2pt
\ligne{\hfill\uu{2 locomotives}\hfill}
\vskip-1pt
\iii {B}{BBWWBBBBWBB}{B}
\iii {B}{BWBWBBBBWBB}{B}
\iii {B}{BWWBBBBBWBB}{B}
}
\hfill
\vtop{\hsize 76pt
\ligne{\hfill\uu{around E}\hfill}
\vskip-3pt
\iii {W}{BBWWWBWWWWW}{W}
\iii {B}{BBWWWBWWWWW}{W}
\iii {W}{BWWBWBWWWWW}{W}
\vskip 2pt
\ligne{\hfill\uu{2 locomotives}\hfill}
\vskip-1pt
\iii {B}{BWWWBBWWWWW}{W}
\iii {W}{BBWWBBWWWWW}{W} 
\iii {W}{BWWWBBWWWWW}{W}
}
\hfill
\vtop{\hsize 76pt
\vskip 3pt
\iii {W}{BWWBWBBWWWW}{W}
\vskip 3pt
\iii {B}{BBWBWWBWWWW}{W}
\iii {W}{BBBWBWBWWWW}{W} 
}
\hfill
}
}

\vskip 5pt
\vtop{\decal=3.5pt
\begin{tab}\label{schedule_selects}
\leurre
The scheduling of the crossing of the selector by a single locomotive.
\end{tab}
\vspace{-9pt}
\ligne{\hfill
\vtop{\footnotesize\tt\hsize=320pt
\ligne{\prerangee {} {i$_1$} i B A {o$_r$} \prerangee {o$_r^1$} {o$_r^2$} 
{o$_\ell$} {o$_\ell^1$} {o$_\ell^2$} D 
\prerangeeii C E 
\hfill}
\ligne{\prerangee {1} {\ennoir {13}} {63} {23} {55} {11}  \prerangee {11} {17} {29} {23} {29} {\ennoir {76}}  
\prerangeeii {\ennoir {69}} {\ennoir {83}} \hfill}
\vskip-2pt
\ligne{\prerangee {2} {14} {\ennoir {64}} {24} {55} {11}  \prerangee {11} {17} {29} {23} {29} {\ennoir {76}}  
\prerangeeii {\ennoir {70}} {\ennoir {83}} \hfill}
\vskip-2pt
\ligne{\prerangee {3} {11} {65} {\ennoir {25}} {56} {11}  \prerangee {11} {17} {29} {23} {29} {\ennoir {77}}  
\prerangeeii {\ennoir {71}} {\ennoir {83}} \hfill}
\vskip-2pt
\ligne{\prerangee {4} {11} {62} {26} {\ennoir {57}} {12}  \prerangee {11} {17} {30} {23} {29} {\ennoir {78}}  
\prerangeeii {\ennoir {22}} {\ennoir {83}} \hfill}
\vskip-2pt
\ligne{\prerangee {5} {11} {66} {89} {58} {\ennoir {13}}  \prerangee {12} {17} {\ennoir {96}} {24} {29} {\ennoir {79}}  
\prerangeeii {72} {\ennoir {84}} \hfill}
\vskip-2pt
\ligne{\prerangee {6} {11} {62} {23} {55} {14}  \prerangee {\ennoir {13}} {18} {97} {\ennoir {90}} {46} {\ennoir {80}}  
\prerangeeii {\ennoir {73}} {85} \hfill}
\vskip-2pt
\ligne{\prerangee {7} {11} {62} {23} {55} {11}  \prerangee {14} {\ennoir {19}} {29} {23} {29} {\ennoir {77}}  
\prerangeeii {\ennoir {69}} {\ennoir {83}} \hfill}
\vskip-2pt
\ligne{\prerangee {8} {11} {62} {23} {55} {11}  \prerangee {11} {17} {29} {23} {29} {\ennoir {76}}  
\prerangeeii {\ennoir {69}} {\ennoir {83}} \hfill}
}
\hfill}
}
\vskip 15pt

Figures~\ref{selectD_mov} and~\ref{selectE_mov} can be viewed as different zooms with respect 
to Figure~\ref{select11} of
Sub-section~\ref{implements}. Tables~\ref{schedule_selects} and~\ref{schedule_selectd} allow
us to follow how the rules are applied to the different cells constituting the selector.

Let us make the description more precise. The arrival by \uu i$_1$ makes use of rules from
Table~\ref{motion} only. Now, \uu i is not an ordinary element of the tracks. It has a specific 
configuration which is illustrated in Figure~\ref{selectD_mov}. Rules~62, 63, 64 and~65 and
also 67 and~68 play the role of motion rules for the cell~\uu i. Rule~66 is used when the 
cell~\uu C is white. Then, the locomotive(s) arrive(s) at~\uu B. This cell is alike an element
of the track with this restriction that two of its milestones may turn white at one moment exactly
and only for that instant. So that the rules of Table~\ref{motion} apply except when \uu C 
or~\uu D is white.
In the first case, rule~89 is applied. In the other one, it is rule~91. The cell~\uu A is ruled
by rules~55 up to~60. Rule~55 manages the idle configuration. Rule~56 make the locomotive enter
the cell and rule~57 kills it at the next time. Rule~58 witnesses that each exit, the one close
to~\uu C, the other close to~\uu D, is occupied by a locomotive. When there are two locomotives,
rules~59 and~60 are applied instead of rules~57 and~58. Rule~61 is used only for the fork
which has exactly the same neighbourhood as an idle cell~\uu A.

\vtop{
\vspace{-5pt}
\ligne{\hfill\hskip 15pt
\includegraphics[scale=0.3]{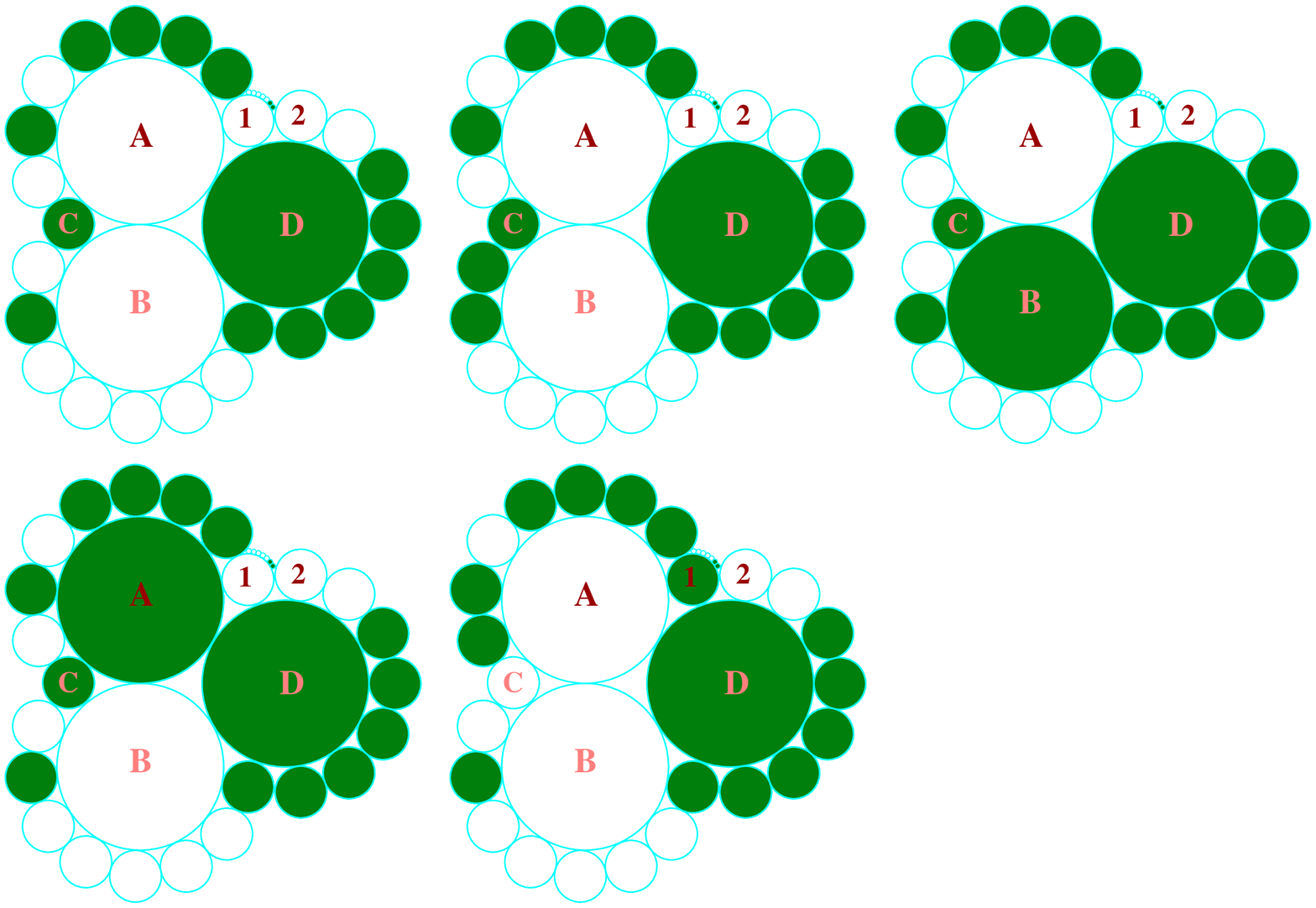}
\hfill}
\vspace{-5pt}
\ligne{\hfill\hskip 15pt
\includegraphics[scale=0.3]{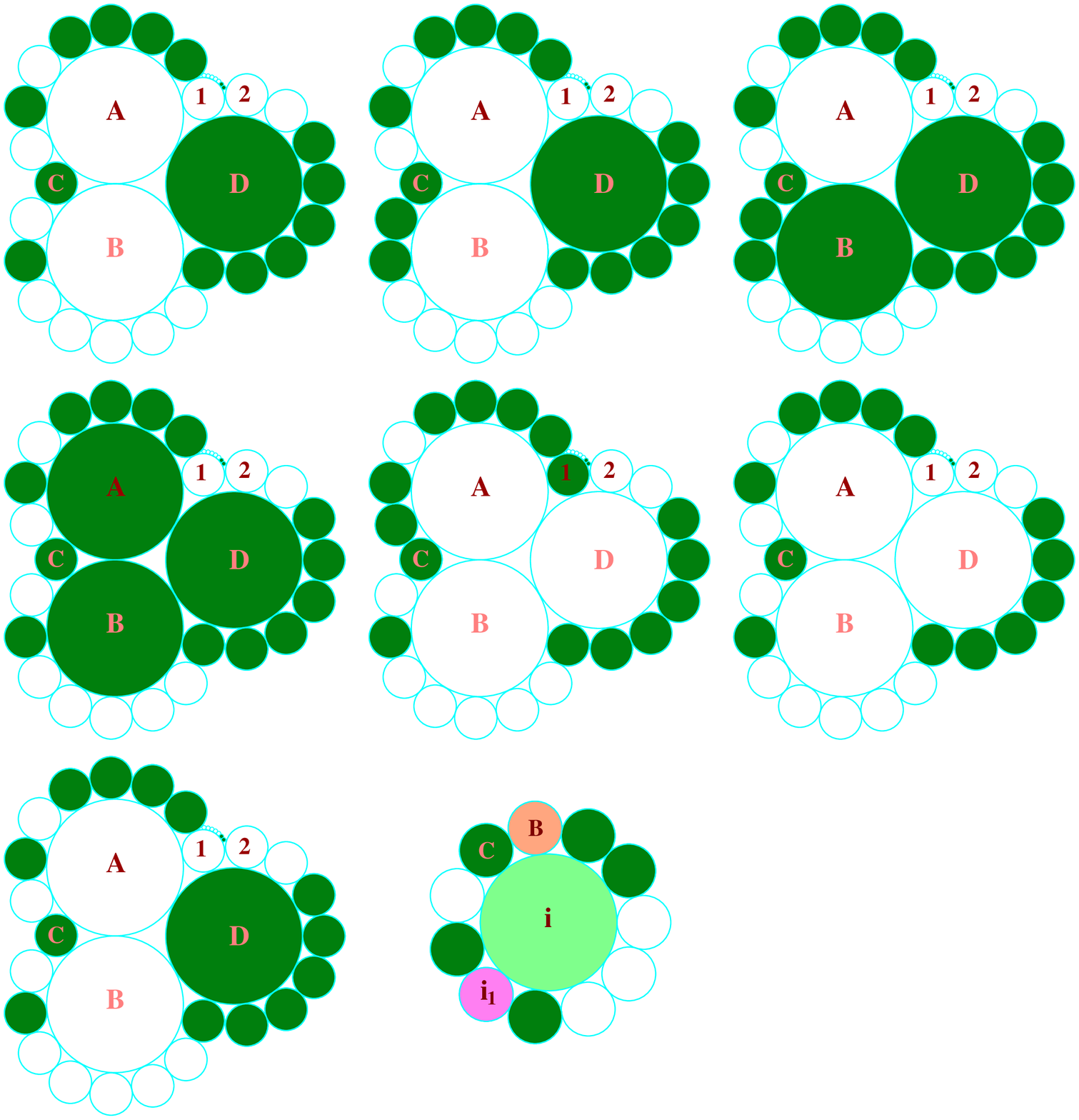}
\hfill}
\begin{fig}\label{selectD_mov}
\leurre
The motion of the locomotive through the selector of a round-about. Focus on the cells~$A$,
$B$ and~$D$. On the last line of the figure, \uu i and its neighbourhood.
\end{fig}
}

   We can notice that the cells~\uu C and~\uu D are applied very different rules depending on the
number of locomotives arriving at the selector. Rules~76 up to~80 are used when there is a single 
locomotive, witnessing the passage through the cells of the track which are in contact with~\uu D.
Rule~81 detects that two locomotives arrived and turn the state of the cell~\uu D to white, while
rule~82 restores the black state at the next tick of the clock. Similarly, rules~69 up to~76
are used the cell~\uu C. It should be noticed that when there is a single locomotive in~\uu A,
the configuration of the neighbours is that of a cell of the track in between two pivots when
a locomotive is in the cell: this is why rule~22 is used to make the cell turn to white.
When there are two locomotives, rules74 up to~76 are used. Note the rules for~\uu E: 
rules~83 up to~85 when there is a single locomotive and rules~86 up to~88 when there are two of 
them.

\vskip 10pt
\vtop{
\ligne{\hfill\hskip 15pt
\includegraphics[scale=0.25]{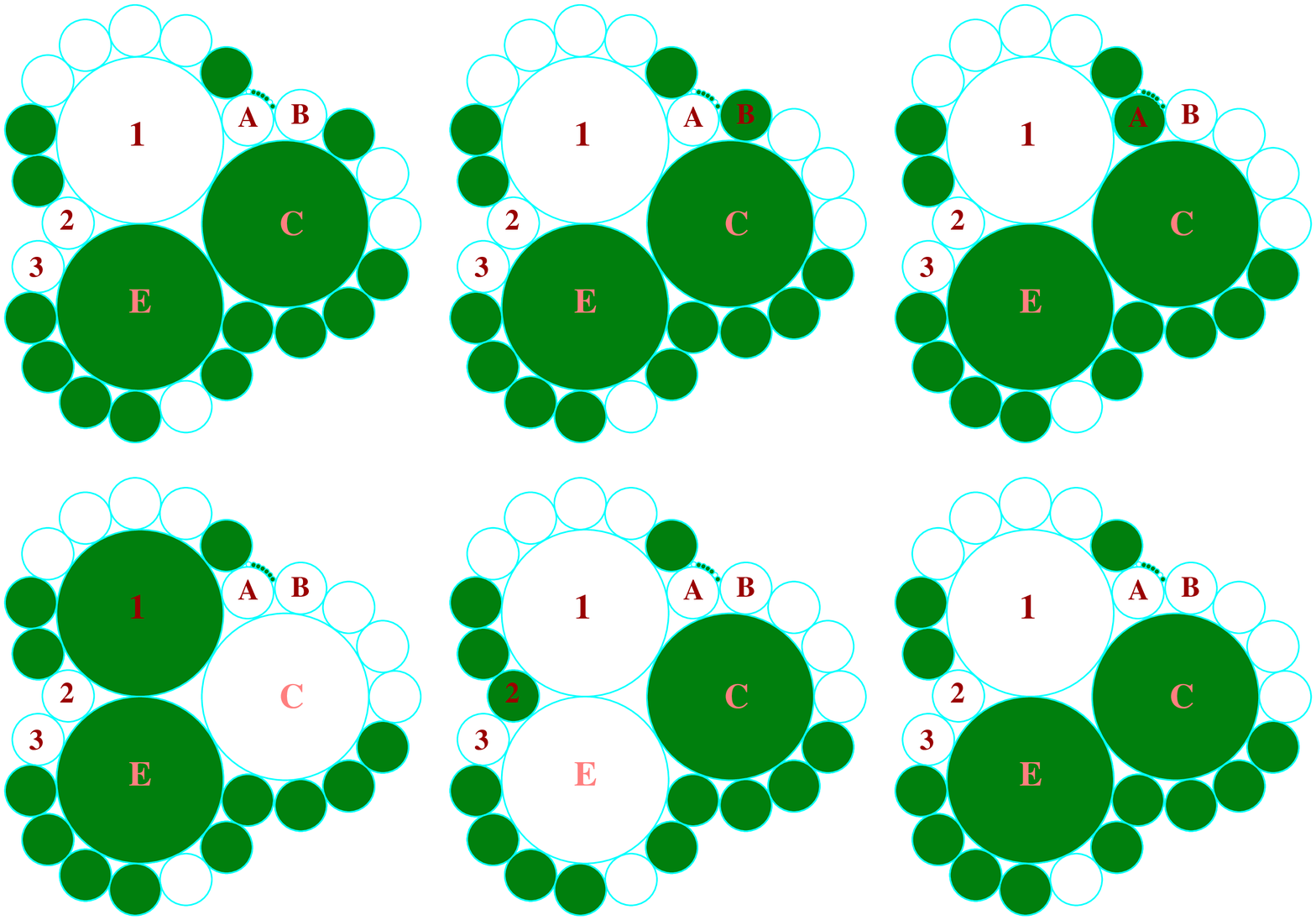}
\hfill}
\vspace{-5pt}
\ligne{\hfill\hskip 15pt
\includegraphics[scale=0.25]{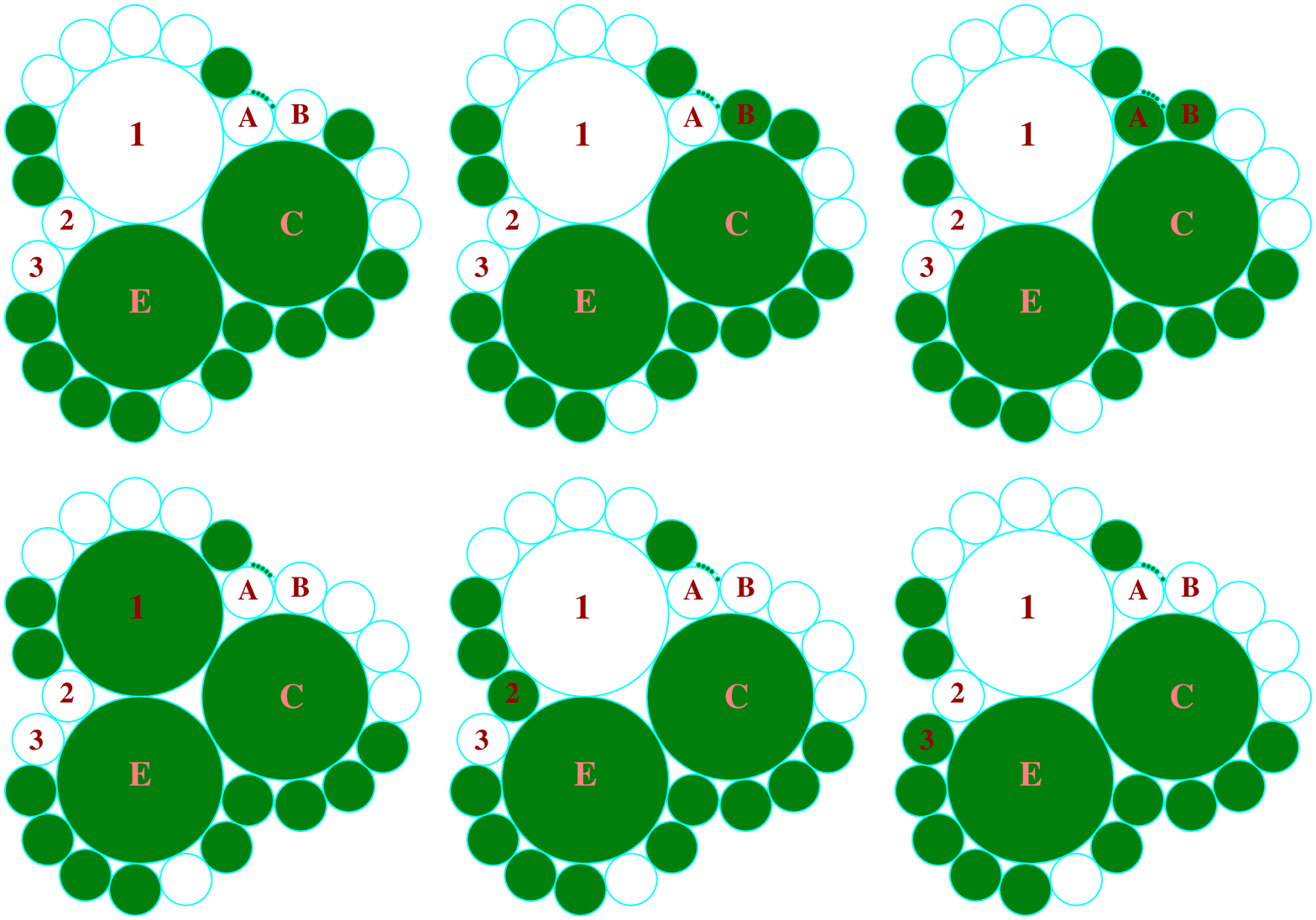}
\hfill}
\begin{fig}\label{selectE_mov}
\leurre
The motion of the locomotive through the selector of a round-about. Focus on the cells~$C$
and~$E$.
\end{fig}
}
\vskip 5pt
   The remaining rules of the table appear in the destruction of the superfluous locomotive:
rule~89 for~\uu B when~\uu C, one of its milestones, momentarily vanishes. Rule~90 kills the 
locomotive which is at~\uu o$_\ell^1$ as the cell~\uu E is white at this instant. Rule~91 is 
used for~\uu B when~\uu D, also one of its milestones, is white for two ticks of the clock.
Rule~92 is used by~\uu o$_r$ as its black but its milestone~\uu D is now white: and so, for the
newt time when \uu D is still white, rule~94 applies. Rules 93 and~94 are used for \uu o$_r^1$
which, in this case remains white but has a missing milestone: \uu D. For a similar reason,
rule~95 is used for \uu o$_r^2$ as for it, \uu D is at another milestone. Eventually, rule~96 
is used by~\uu o$_\ell$ when \uu C is white and rule~97 is used at the next time, \uu E being
then white.

\vtop{\decal=3.5pt
\begin{tab}\label{schedule_selectd}
\leurre
The scheduling of the crossing of the selector by two contiguous locomotives.
\end{tab}
\vspace{-9pt}
\ligne{\hfill
\vtop{\footnotesize\tt\hsize=320pt
\ligne{\prerangee {} {i$_1$} i B A {o$_r$} \prerangee {o$_r^1$} {o$_r^2$} 
{o$_\ell$} {o$_\ell^1$} {o$_\ell^2$} D 
\prerangeeii C E 
\hfill}
\ligne{\prerangee {1} {\ennoir {16}} {\ennoir {67}} {24} {55} {11}  \prerangee {11} {17} {29} {23} {29} {\ennoir {76}}  
\prerangeeii {\ennoir {70}} {\ennoir {83}} \hfill}
\vskip-2pt
\ligne{\prerangee {2} {14} {\ennoir {68}} {\ennoir {27}} {56} {11}  \prerangee {11} {17} {29} {23} {29} {\ennoir {77}}  
\prerangeeii {\ennoir {74}} {\ennoir {83}} \hfill}
\vskip-2pt
\ligne{\prerangee {3} {11} {65} {\ennoir {28}} {\ennoir {59}} {12}  \prerangee {11} {17} {30} {23} {29} {\ennoir {81}}  
\prerangeeii {\ennoir {75}} {\ennoir {83}} \hfill}
\vskip-2pt
\ligne{\prerangee {4} {11} {62} {91} {60} {\ennoir {92}}  \prerangee {93} {95} {\ennoir {31}} {24} {29} {82}  
\prerangeeii {\ennoir {76}} {\ennoir {86}} \hfill}
\vskip-2pt
\ligne{\prerangee {5} {11} {62} {91} {65} {94}  \prerangee {94} {95} {32} {\ennoir {25}} {30} {72}  
\prerangeeii {\ennoir {69}} {\ennoir {87}} \hfill}
\vskip-2pt
\ligne{\prerangee {6} {11} {62} {23} {55} {11}  \prerangee {11} {17} {29} {26} {\ennoir {31}} {\ennoir {76}}  
\prerangeeii {\ennoir {69}} {\ennoir {88}} \hfill}
\vskip-2pt
\ligne{\prerangee {7} {11} {62} {23} {55} {11}  \prerangee {11} {17} {29} {23} {29} {\ennoir {76}}  
\prerangeeii {\ennoir {69}} {\ennoir {83}} \hfill}
\vskip-2pt
\ligne{\prerangee {8} {11} {62} {23} {55} {11}  \prerangee {11} {17} {29} {23} {29} {\ennoir {76}}  
\prerangeeii {\ennoir {69}} {\ennoir {83}} \hfill}
}
\hfill}
}
\vskip 10pt

\subsection{The active switches}

    From Section~\ref{scenar}, we know that we have two active switches: the flip-flop
and the part of the memory switch which is actively crossed by the locomotive. For these
switches, we use two common structures: the fork and the controller. Figures~\ref{flipflop}
and~\ref{memoactive} from Subsection~\ref{genpatt} illustrate how these patterns are assembled
in each case.

\subsubsection*{Fork}

\vskip-10pt
\vtop{
\ligne{\hfill\hskip 15pt
\includegraphics[scale=0.25]{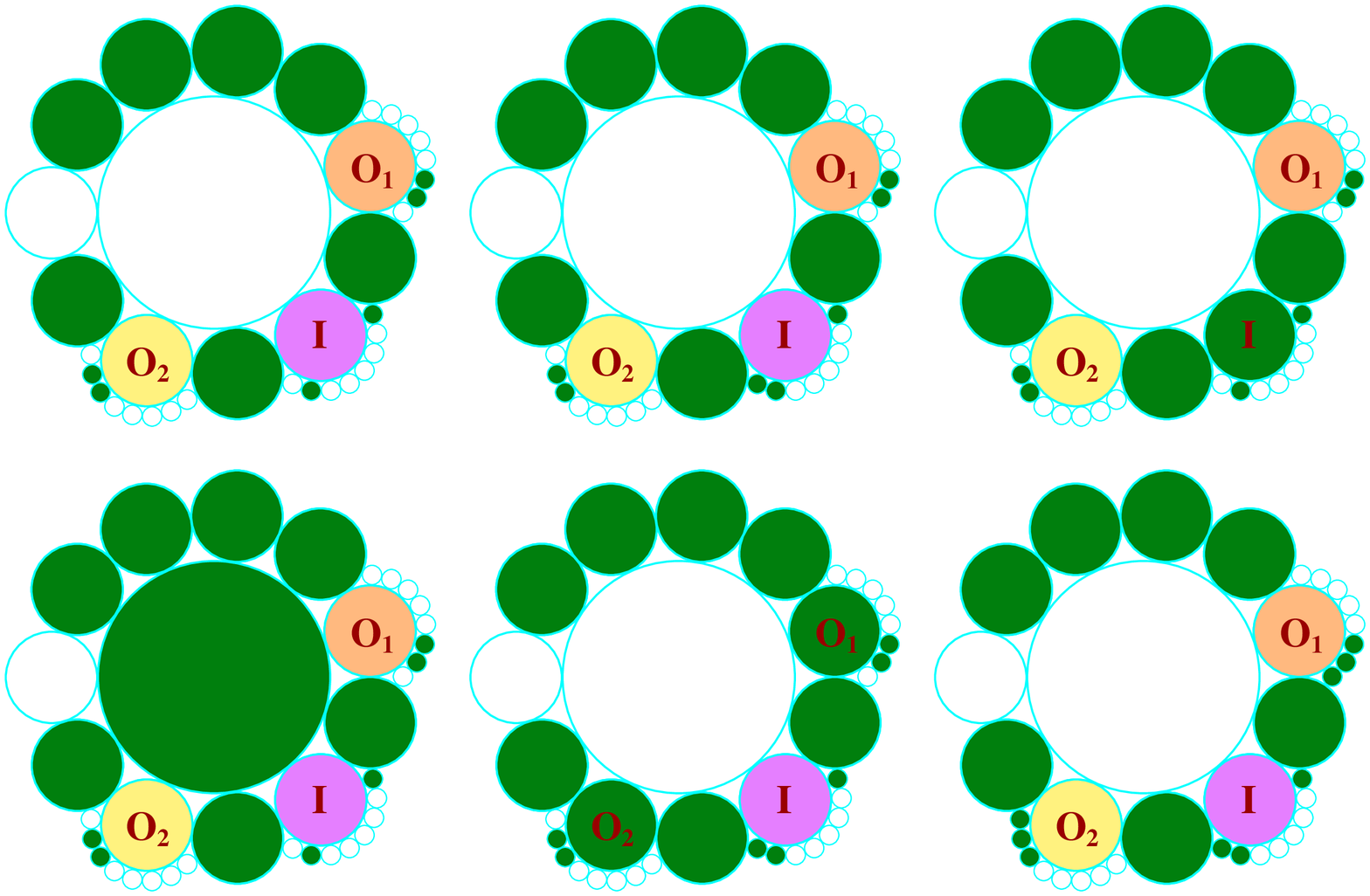}
\hfill}
\begin{fig}\label{fork11_mov}
\leurre
The motion of the locomotive through the fork in the switches. 
\end{fig}
}

   The idle configuration of the fork is the same as that of the cell~\uu A in the selector.
This can be checked on Figure~\ref{fork11_mov}. The centre of the fork makes use of the 
rules~55, 56 and 57 from those used by the cell~\uu A. As already mentioned it also makes use
of rule~61. The slight differences with the rules applied to the cell~\uu A of the selector
come from the fact that in the neighbourhood of the centre of the fork, the black cells are
invariant.

\vtop{\decal=4pt
\begin{tab}\label{schedule_fork}
\leurre
The scheduling of the crossing of the fork by a single locomotive.
\end{tab}
\vspace{-9pt}
\ligne{\hskip 80pt
\vtop{\footnotesize\tt\hsize=230pt
\ligne{\prerangee {} {i$_1$} i t o {o$_r$} \prerangeeii {o$_\ell$} {o$_\ell^1$} 
\hfill}
\ligne{\prerangee {1} {\ennoir {19}} {24} {55} {11} {17}  \prerangeeii {11} {17} \hfill}
\vskip-2pt
\ligne{\prerangee {2} {20} {\ennoir {25}} {56} {11} {17}  \prerangeeii {11} {17} \hfill}
\vskip-2pt
\ligne{\prerangee {3} {17} {26} {\ennoir {57}} {12} {17}  \prerangeeii {12} {17} \hfill}
\vskip-2pt
\ligne{\prerangee {4} {17} {23} {61} {\ennoir {13}} {18}  \prerangeeii {\ennoir {13}} {18} \hfill}
\vskip-2pt
\ligne{\prerangee {5} {17} {23} {55} {14} {\ennoir {19}}  \prerangeeii {14} {\ennoir {19}} \hfill}
\vskip-2pt
\ligne{\prerangee {6} {17} {23} {55} {11} {17}  \prerangeeii {11} {17} \hfill}
}
\hfill}
}
\vskip 15pt
As will be seen further again, the fork will also be used in the passive memory switch as 
already shown in Figure~\ref{memopassive} from Subsection~\ref{genpatt}. 

\vskip-5pt
\subsubsection*{Controller}

    For checking the controller of the active switches, we just need Figure~\ref{switchactive}
and the explanations of Subsection~\ref{genpatt}. The rules are given by 
Table~\ref{control_rules}.

\vskip-10pt
\vtop{
\begin{tab}\label{control_rules}
\leurre
Rules for the controller used by the active switches.
\end{tab}
\vspace{-9pt}
\ligne{\hfill\footnotesize\tt
\hbox to 76pt{\hfill the cell {\bf t}\hfill}
\hskip 15pt
\hbox to 167pt{\hfill the cell~{\bf c}\hfill}
\hfill
}
\ligne{\hfill\footnotesize\tt
\vtop{\hsize=76pt
\ligne{\uu{\hskip 19pt C\hskip 10pt o  i}\hfill}
\vskip-3pt
\iii {W}{BBBWBBBWWWW}{W}
}\hskip 15pt
\vtop{\hsize=76pt
\ligne{\uu{\hskip 47pt t\hskip 18pt s}\hfill}
\vskip-3pt
\hskip 4.5pt\iii {W}{BBBBBWWBBBW}{W}
\iii {B}{BBBBBWWBBBW}{B}
}
\hskip 15pt
\vtop{\hsize=76pt
\ligne{\uu{\hskip 47pt t\hskip 18pt s}\hfill}
\vskip-3pt
\iii {W}{BBBBBWWBBBB}{B}
\iii {B}{BBBBBWWBBBB}{W}
}
\hfill
}
}
\vskip 15pt

   From the right-hand side of Figure~\ref{switchactive}, when the cell~\uu C of the controller
is white, the cell~\uu t looks like a cell of the tracks: its neighbourhood is exactly
one of the neighbourhoods for cells which are elements of the tracks. Note the use of rule~61
for the cell~\uu C when the controller is white, leaving the locomotive cross the controller,
see the lower half of Table~\ref{schedule_controlv}.

\def\yyy{\hskip-3.5pt}
\vskip-15pt
\vtop{\decal=4pt
\begin{tab}\label{schedule_controlv}
\leurre
The scheduling of the passage of the locomotive through the cell\uu t, depending on whether
the controller is black, upper table, or white, lower table.
\end{tab}
\vspace{-9pt}
\ligne{\hskip 60pt
\vtop{\footnotesize\tt\hsize=230pt
\ligne{\prerangee {} {i$_1$} i t o {o$_r$} \prerangeeiv {o$_r^1$} C {s} {s$_1$}
\hfill}
\ligne{\prerangee {1} {\ennoir {13}} {12} {97} {11} {11}  \prerangeeiv {11} {\yyy\ennoir {100}} 
{26} {11} \hfill}
\vskip-2pt
\ligne{\prerangee {2} {14} {\ennoir {13}} {98} {11} {11}  \prerangeeiv {11} {\yyy\ennoir {100}} 
{26} {11} \hfill}
\vskip-2pt
\ligne{\prerangee {3} {11} {11} {97} {11} {11}  \prerangeeiv {11} {\yyy\ennoir {100}} {26} {11} 
\hfill}
\vskip-2pt
\ligne{\prerangee {4} {11} {11} {97} {11} {11}  \prerangeeiv {11} {\yyy\ennoir {100}} {26} {11} 
\hfill}
}
\hfill}
\vskip 5pt
\ligne{\hskip 60pt
\vtop{\footnotesize\tt\hsize=230pt
\ligne{\prerangee {} {i$_1$} i t o {o$_r$} \prerangeeiv {o$_r^1$} C {s} {s$_1$}
\hfill}
\ligne{\prerangee {1} {\ennoir {13}} {12} {23} {11} {11}  \prerangeeiv {11} {99} {23} {11} \hfill}
\vskip-2pt
\ligne{\prerangee {2} {14} {\ennoir {13}} {24} {11} {11}  \prerangeeiv {11} {99} {23} {11} \hfill}
\vskip-2pt
\ligne{\prerangee {3} {11} {14} {\ennoir {25}} {12} {11}  \prerangeeiv {11} {61} {23} {11} \hfill}
\vskip-2pt
\ligne{\prerangee {4} {11} {11} {26} {\ennoir {13}} {12}  \prerangeeiv {11} {99} {23} {11} \hfill}
\vskip-2pt
\ligne{\prerangee {5} {11} {11} {23} {14} {\ennoir {13}}  \prerangeeiv {12} {99} {23} {11} \hfill}
\vskip-2pt
\ligne{\prerangee {6} {11} {11} {23} {11} {14}  \prerangeeiv {\ennoir {13}} {99} {23} {11} \hfill}
\vskip-2pt
\ligne{\prerangee {7} {11} {11} {23} {11} {11}  \prerangeeiv {11} {99} {23} {11} \hfill}
}
\hfill}
}
\vskip 10pt
   When the cell~\uu C is black, rule~97 manages the idle configuration of~\uu t and rule~98 
prevents a locomotive to enter~\uu t when \uu C is black. The two right-hand side columns of 
Table~\ref{control_rules} handle the behaviour of the cell~\uu C.

\def\zz{\hskip-2pt}
\vtop{\decal=4pt
\vspace{-10pt}
\begin{tab}\label{schedule_controlch}
\leurre
The scheduling of the change in the controller.
Upper table: from unlocked to locked. Lower table: from locked to unlocked.
\end{tab}
\vspace{-9pt}
\ligne{\hskip 60pt
\vtop{\footnotesize\tt\hsize=230pt
\ligne{\prerangee {} {i$_1$} i t o {o$_r$} \prerangeeiv {o$_r^1$} C {s} {s$_1$}
\hfill}
\ligne{\prerangee {1} {11} {11} {23} {11} {11}  \prerangeeiv {11} {99} {24} {\ennoir {13}} \hfill}
\vskip-2pt
\ligne{\prerangee {2} {11} {11} {23} {11} {11}  \prerangeeiv {11} {101} {\ennoir {25}} {14} \hfill}
\vskip-2pt
\ligne{\prerangee {3} {11} {11} {97} {11} {11}  \prerangeeiv {11} {\yyy\ennoir {100}} {26} {11} 
\hfill}
\vskip-2pt
\ligne{\prerangee {4} {11} {11} {97} {11} {11}  \prerangeeiv {11} {\yyy\ennoir {100}} {26} {11} 
\hfill}
}
\hfill}
\vskip 5pt
\ligne{\hskip 60pt
\vtop{\footnotesize\tt\hsize=230pt
\ligne{\prerangee {} {i$_1$} i t o {o$_r$} \prerangeeiv {o$_r^1$} C {s} {s$_1$}
\hfill}
\ligne{\prerangee {1} {11} {11} {97} {11} {11}  \prerangeeiv {11} {\yyy\ennoir {100}} {72} {\ennoir {13}} \hfill}
\vskip-2pt
\ligne{\prerangee {2} {11} {11} {97} {11} {11}  \prerangeeiv {11} {\yyy\ennoir {102}} {\ennoir {28}} {14} \hfill}
\vskip-2pt
\ligne{\prerangee {3} {11} {11} {23} {11} {11}  \prerangeeiv {11} {99} {23} {11} \hfill}
\vskip-2pt
\ligne{\prerangee {4} {11} {11} {23} {11} {11}  \prerangeeiv {11} {99} {23} {11} \hfill}
}
\hfill}
}
\vskip 15pt
Rules~99 and~100 manage the idle configuration in which
\uu C may be either white or black but permanently in the same state. 
Rules~101 and~102 handle the change of state of~\uu C which is triggered by the arrival of
a locomotive through its neighbour~11. Note that both rules have exactly the same context
as in both cases, the neighbourhood of~\uu C is the same.

\subsubsection{The passive memory switch}

   From Subsection~\ref{genpatt}, we know that the new pattern involved in the passive memory
switch is the structure we called the sensor. The structure is illustrated by 
Figure~\ref{sensor11}.

\subsubsection*{Sensor}

   As in the case of the controller, if the cell~\uu C is white, the cell~\uu t
behaves as a cell of the track as it has one of the defined neighbourhoods for the elements
of the tracks. However, and it is one of the differences with what happens in the controller,
when \uu C is black, the locomotive must nevertheless cross the cell~\uu t. So that in this
case, we have new motion rules given by rules~103 up to~106, that one included.

This time, the cell~\uu C has two auxiliary cells~\uu S and~\uu E.
Rules~114 and~115 handle the case of the idle configuration for~\uu S, no matter which 
is the state of~\uu C. When the cell~\uu C is white, if a locomotive runs through~\uu t, 
\uu C remains white as this is the case of a passage through the selected track: this is checked 
by rule~107. When \uu C is black, this means that the track is not selected. The passage of 
the locomotive requires that \uu C changes to white and that it will keep the state white until 
a signal of a change comes through the cell~\uu E. The change from black to white for~\uu C 
is controlled by rule~108. But at the same time, \uu S must flash from white to black and back
to white through rules~56 and~68: look at Table~\ref{schedule_sensor}.
When \uu S has flashed, note the locomotive moving from~\uu v$_S$ and~\uu v$_S^1$: this locomotive
will reach the other sensor of the switch to there make the cell~\uu E flash in order to turn the
cell~\uu C from white to black.

\vtop{
\vspace{-15pt}
\begin{tab}\label{sensor_rules}
\leurre
Rules for the sensor used by the passive memory switch.
\end{tab}
\vspace{-9pt}
\ligne{\hfill\footnotesize\tt
\hfill \hbox to 76pt{\hfill cell {\bf t}\hfill}\hskip 7pt
\hbox to 81pt{\hfill cell {\bf c}\hfill}\hskip 7pt
\hbox to 81pt{\hfill cell {\bf e}\hfill}\hskip 7pt
\hbox to 81pt{\hfill cell {\bf s}\hfill}
\hfill}
\ligne{\hfill\footnotesize\tt
\vtop{\hsize=76pt
\ligne{\uu{\hskip 24pt C\hskip 10pt o\hskip 9pt i}\hfill}
\vskip-3pt
\iii {W}{BBBWBBWBWWW}{W}
\iii {W}{BBBWBBBBWWW}{B}
\iii {B}{BBBWBBWBWWW}{W}
\iii {W}{WBBBBBWBWWB}{W}
}
\hskip 7pt
\vtop{\hsize=81pt
\ligne{\uu{\hskip 53pt t\xxx S \hskip-2pt E}\hfill}
\vskip-3pt
\iii {W}{BBBBBBBWBWW}{W}
\iii {B}{BBBBBBBWBWW}{W}
\iii {W}{BBBBBBWWBBW}{B}
}
\hskip 7pt
\vtop{\hsize=81pt
\ligne{\uu{\hskip 24pt C\hskip 18pt i}\hfill}
\vskip-3pt
\iii {W}{WBWBBWBBBWW}{W}
\iii {W}{BBWBBWBBBWW}{W}
\iii {W}{WBWBBBBBBWW}{B}
\iii {B}{WBWBBWBBBWW}{W}
}
\hskip 7pt
\vtop{\hsize=81pt
\ligne{\uu{\hskip 24pt C\xxx t\hskip 32pt o}\hfill}
\vskip-3pt
\iii {W}{BWWBBBBWBWB}{W}
\iii {W}{WWWBBBBWBBB}{W}
}
\hfill}
\ifnum 1=0 {
} \fi
\vskip10pt
}
\ifnum 1=0 {
} \fi

Let us look at that: the auxiliary locomotive arrives to~\uu v$_E^1$ and then to~\uu v$_E$.
As can be seen in Table~\ref{schedule_sensor}, rule~112 makes~\uu E turn from white to black
and then rule~113 makes in turn back to white. But when~\uu C can see that~\uu E is black,
it turns from white to black thanks to rule~109. Note that for~\uu E, rules~110 and~111 manage
its idle configuration whatever the state of~\uu C.

\def\zz{\hskip-4pt}
\vtop{\decal=4pt
\vspace{-10pt}
\begin{tab}\label{schedule_sensor}
\leurre
The scheduling of the passage of the locomotive controlled by the sensor.
\end{tab}
\vspace{-9pt}
\ligne{\hskip 60pt
\vtop{\footnotesize\tt\hsize=300pt
\ligne{\prerangee {} {i$_1$} i t o {o$_r$} \prerangee C E  S {v$_E$} {v$_E^1$} {v$_S$}
\prerangeei {v$_E^1$} \hfill}
\ligne{\prerangee {1} {\ennoir {13}} {12} {103} {11} {11}  \prerangee {\ennoir {79}} {111} {114} {11} {11} {11}  
\prerangeei {11} \hfill}
\vskip-2pt
\ligne{\prerangee {2} {14} {\ennoir {13}} {104} {11} {11}  \prerangee {\ennoir {79}} {111} {114} {11} {11} {11}  
\prerangeei {11} \hfill}
\vskip-2pt
\ligne{\prerangee {3} {11} {14} {\yyy\ennoir {105}} {12} {11}  \prerangee {\yyy\ennoir {108}} 
{111} {56} {11} {11} {11}  
\prerangeei {11} \hfill}
\vskip-2pt
\ligne{\prerangee {4} {11} {11} {106} {\ennoir {13}} {12}  \prerangee {58} {110} {\ennoir {68}} {11} {11} {12}  
\prerangeei {11} \hfill}
\vskip-2pt
\ligne{\prerangee {5} {11} {11} {29} {14} {\ennoir {13}}  \prerangee {82} {110} {115} {11} {11} {\ennoir {13}}  
\prerangeei {12} \hfill}
\vskip-2pt
\ligne{\prerangee {6} {11} {11} {29} {11} {11}  \prerangee {82} {110} {65} {11} {11} {14}  
\prerangeei {\ennoir {13}} \hfill}
}
\hfill}
\vskip 7pt
\ligne{\hskip 60pt
\vtop{\footnotesize\tt\hsize=300pt
\ligne{\prerangee {} {i$_1$} i t o {o$_r$} \prerangee C E  S {v$_E$} {v$_E^1$} {v$_S$}
\prerangeei {v$_E^1$} \hfill}
\ligne{\prerangee {1} {\ennoir {13}} {12} {29} {11} {11}  \prerangee {82} {110} {65} {11} {11} {11}  
\prerangeei {11} \hfill}
\vskip-2pt
\ligne{\prerangee {2} {14} {\ennoir {13}} {30} {11} {11}  \prerangee {82} {110} {65} {11} {11} {11}  
\prerangeei {11} \hfill}
\vskip-2pt
\ligne{\prerangee {3} {11} {14} {\ennoir {31}} {12} {11}  \prerangee {107} {110} {55} {11} {11} {11}  
\prerangeei {11} \hfill}
\vskip-2pt
\ligne{\prerangee {4} {11} {11} {32} {\ennoir {13}} {12}  \prerangee {82} {110} {65} {11} {11} {11}  
\prerangeei {11} \hfill}
\vskip-2pt
\ligne{\prerangee {5} {11} {11} {29} {14} {\ennoir {13}}  \prerangee {82} {110} {65} {11} {11} {11}  
\prerangeei {11} \hfill}
\vskip-2pt
\ligne{\prerangee {6} {11} {11} {29} {11} {11}  \prerangee {82} {110} {65} {11} {11} {11}  
\prerangeei {11} \hfill}
}
\hfill}
\vskip 7pt
\ligne{\hskip 60pt
\vtop{\footnotesize\tt\hsize=300pt
\ligne{\prerangee {} {i$_1$} i t o {o$_r$} \prerangee C E  S {v$_E$} {v$_E^1$} {v$_S$}
\prerangeei {v$_E^1$} \hfill}
\ligne{\prerangee {1} {11} {11} {29} {11} {11}  \prerangee {82} {110} {65} {12} {\ennoir {13}} {11}  
\prerangeei {11} \hfill}
\vskip-2pt
\ligne{\prerangee {2} {11} {11} {29} {11} {11}  \prerangee {82} {112} {65} {\ennoir {13}} {14} {11}  
\prerangeei {11} \hfill}
\vskip-2pt
\ligne{\prerangee {3} {11} {11} {29} {11} {11}  \prerangee {109} {\yyy\ennoir {113}} {65} {14} 
{11} {11}  
\prerangeei {11} \hfill}
\vskip-2pt
\ligne{\prerangee {4} {11} {11} {103} {11} {11}  \prerangee {\ennoir {79}} {111} {114} {11} {11} {11}  
\prerangeei {11} \hfill}
\vskip-2pt
\ligne{\prerangee {5} {11} {11} {103} {11} {11}  \prerangee {\ennoir {79}} {111} {114} {11} {11} {11}  
\prerangeei {11} \hfill}
\vskip-2pt
\ligne{\prerangee {6} {11} {11} {103} {11} {11}  \prerangee {\ennoir {79}} {111} {114} {11} {11} {11}  
\prerangeei {11} \hfill}
}
\hfill}
}
\vskip 10pt

   With the help of the figures, of the tables for the rules and of the schedule tables we have
proved the following result:

\vskip-5pt
\vtop{
\begin{thm}
There is a weakly universal cellular automaton on the tiling $\{11,3\}$ which is planar and 
rotation invariant.
\end{thm}
}

   We remind the reader that with {\it planar}, we mean that the set of cells which are crossed
by the locomotive(s) contains infinitely many cycles.

\subsection{Concluding remarks}

   It can be asked whether it is possible or not to lower the number of neighbours in order
to get a planar rotation invariant weakly universal cellular automaton in a tiling $\{p,3\}$
or $\{p,4\}$.

   A first remark on the number of rules. 

   The tables we have displayed for the rules indicate 115 of them. However, there might be more
rules: this depends on what we call the program of the cellular automaton. We should notice
that these 115~rules are rotationally independent: none of them can be obtained from another
one by a circular permutation on the neighbour's states. Accordingly, if we consider
that the program of a rotation invariant cellular automaton should contain all rotated forms
of the same rule, there should be much more rules: roughly 115$\times$11. However, this is 
not exactly true. If we consider the motion rules for the tracks, then it is likely that
any rotated form will be met in the construction of the tracks needed for the simulation of a
register machine. However, for the cells concerning a structure, this may be not the case.
It may be arranged that all configurations make use of the same rules which are the ones used
in the tables of the paper, or that we have only to take two or three rotated forms of the same 
rules but not all of them.

   Table~\ref{biuse} gives us the list of the rules which occur in two different settings. 
As can be checked, for the
time and the cell to which they apply, the configurations of the neighbours and the state
of the cell is the same.

  It should be noticed that the same cell may appear several times in this table. As an example,
the cell~\uu t of the fork appears four times in the table. It appears twice when the fork is idle,
also when the locomotive is about to enter~\uu t and when two locomotives leave the fork. The 
neighbourhood is that of the cell~\uu A of the selector when it is idle, when the locomotive is 
in~\uu i in the selector, when the locomotive is in~\uu A in the selector and for the cell~\uu C
of the controller when the locomotive there is in~\uu t. In rules~57 and~61 which both apply to
the fork and something else,
the context differ by the fact that \uu o$_r$ is white in rule~57 but black in rule~61. 
We can see that the neighbourhood of the cell is rotated in one rule with respect to what it
is in the other rule.

   Note also that rules~79 and~82 have the same context, but
the state of the cell is different. In this case each rule also applies to two different cells.
However, there are several cases of different rules having the same context. This is in particular
the case for motion rules applied to the same cell and also for rules managing the cell~\uu C
in the controller and the cell~\uu C in the sensor.

    It is possible that the rather big number of rules applied in different contexts indicates
that it must be difficult to lower the number of sides of a tile in a tiling $\{p,3\}$ in order
to obtain a weakly universal cellular automaton which would be rotation invariant and which would
be planar. This seems difficult at least using the railway model. I say it might be difficult, 
I cannot say it is impossible.

\newdimen\novlarge\novlarge=110pt
\newdimen\novsouslarge\novsouslarge=75pt
\def\novrangee #1 #2 #3 #4 {
\ligne{\hfill
\hbox to \novlarge{\hfill #1\hfill}
\hbox to \novsouslarge{#2\hfill}
\hbox to \novsouslarge{#3\hfill}
\hbox to \novsouslarge{#4\hfill}
\hfill}
}

\vskip-10pt
\vtop{
\begin{tab}\label{biuse}
\leurre
Table of rules which are applied at list twice, each time in a different context. For each rule,
identified by its number,
the table gives the cell to which it is applied by its name in the structure. The time is also
indicated and, possibly as an index, the case where it appears in its schedule table. Also,
the context of the cell is displayed above the rule.
\end{tab}
\vskip-6pt
\regle=46
\novrangee {\raise 7pt\hbox{\vtop{\hsize=76pt\tt
            \ligne{\footnotesize\hskip 20pt F\hskip 3pt i\hskip 6pt o\hfill}
            \vskip-6pt
            \iii {W}{BBWBWBBWWWW}{W}
            \vskip-2pt
            \ligne{\footnotesize\hskip 29pt o\hskip 4pt E\xxx i\hfill}
           }}
           }
           {{\bf fix:}\hskip 5pt \uu i$_r$, 3} {{\bf sel:}\hskip 5pt \uu o$_\ell^2$, 6$_1$}
           {}
\vskip 6pt
\advance \regle by 8
\novrangee {\raise 7pt\hbox{\vtop{\hsize=76pt\tt
\ligne{\uu{\hskip 19pt C\xxx B\xxx D\xxx o$_r$ \hskip 18pt o$_\ell$}\hfill}
\vskip-6pt
\iii {W}{BWBWBBBBWBW}{W}
\vskip-2pt
\ligne{\uu{\hskip 19pt \ \xxx i\xxx \ \xxx o$_r$ \hskip 18pt o$_\ell$}\hfill}
\vskip-7pt
\ligne{\uu{\hskip 20pt C\xxx t\xxx \hskip 36pt v$_S$}\hfill}
}}
}
          {{\bf sel:}\hskip 5pt \uu A, 0} {{\bf for:}\hskip 5pt \uu t, 0} 
          {{\bf sen:}\hskip 5pt \uu S, 3}
\vskip 6pt

\novrangee {\raise 7pt\hbox{\vtop{\hsize=76pt\tt
\ligne{\uu{\hskip 19pt C\xxx B\xxx D\xxx o$_r$ \hskip 18pt o$_\ell$}\hfill}
\vskip-6pt
\iii {W}{BBBWBBBBWBW}{B}
\vskip-2pt
\ligne{\uu{\hskip 19pt \ \xxx i\xxx \ \xxx o$_r$ \hskip 18pt o$_\ell$}\hfill}
\vskip-7pt
\ligne{\uu{\hskip 20pt C\xxx t\xxx \hskip 36pt v$_S$}\hfill}
}}
}
          {{\bf sel:}\hskip 5pt \uu A, 2$_2$, 3$_1$} {{\bf for}\hskip 5pt \uu t, 2} 
          {{\bf sen:}\hskip 5pt \uu S, 3}

\vskip 6pt
\novrangee {\raise 7pt\hbox{\vtop{\hsize=76pt\tt
\ligne{\uu{\hskip 19pt C\xxx B\xxx D\xxx o$_r$ \hskip 18pt o$_\ell$}\hfill}
\vskip-6pt
\iii {B}{BWBWBBBBWBW}{W}
\vskip-2pt
\ligne{\uu{\hskip 19pt \ \xxx i\xxx \ \xxx o$_r$ \hskip 18pt o$_\ell$}\hfill}
}}
}
          {{\bf sel:}\hskip 5pt \uu A, 4$_1$} {{\bf for:}\hskip 5pt \uu t, 3} {}

\vskip 6pt
\novrangee {\raise 7pt\hbox{\vtop{\hsize=76pt\tt
\ligne{\uu{\hskip 19pt C\xxx B\xxx D\xxx o$_r$ \hskip 18pt o$_\ell$}\hfill}
\vskip-6pt
\iii {W}{WWBBBBBBWBB}{W}
\vskip-2pt
\ligne{\uu{\hskip 53pt t\xxx S \hskip-2pt E}\hfill}
}}
}
          {{\bf sel:}\hskip 5pt \uu A, 5} {{\bf sen:}\hskip 5pt \uu C, 0} {}

\advance \regle by 2
\vskip 6pt
\novrangee {\raise 7pt\hbox{\vtop{\hsize=76pt\tt
\ligne{\uu{\hskip 19pt \ \xxx o$_r$\hskip 23pt o$_\ell$\hskip 1pt i}\hfill}
\vskip-6pt
\iii {W}{BBBBBBWBBBW}{W}
\vskip-2pt
\ligne{\uu{\hskip 43.5pt t\hskip 18pt s}\hfill}
}}
}
          {{\bf for:}\hskip 5pt \uu t, 4} {{\bf con:}\hskip 5pt \uu C, 3} {}

\advance \regle by 3
\vskip 6pt
\novrangee {\raise 7pt\hbox{\vtop{\hsize=76pt\tt
\ligne{\uu{\hskip 28pt B \hskip 10pt i$_1$}\hfill}
\vskip-6pt
\iii {W}{BBBBWBWBWWW}{W}
\vskip-2pt
\ligne{\uu{\hskip 47pt v$_S$C\xxx t}\hfill}
}}
}
          {{\bf sel:}\hskip 5pt \uu i, 3; \uu A, 5} {{\bf sen:}\hskip 5pt \uu S, 0} {}

\advance \regle by 2
\vskip 6pt
\novrangee {\raise 7pt\hbox{\vtop{\hsize=76pt\tt
\ligne{\uu{\hskip 28pt B \hskip 10pt i$_1$}\hfill}
\vskip-6pt
\iii {B}{BBBBWBWBWWW}{W}
\vskip-2pt
\ligne{\uu{\hskip 47pt v$_S$C\xxx t}\hfill}
}}
}
          {{\bf sel:}\hskip 5pt \uu i, 2} {{\bf sen:}\hskip 5pt \uu S, 4} {}

\advance \regle by 3
\vskip 6pt
\novrangee {\raise 7pt\hbox{\vtop{\hsize=76pt\tt
\ligne{\uu{\hskip 19pt AoE \hskip 28pt B}\hfill}
\vskip-6pt
\iii {W}{WBBBBBBWWWW}{B}
\vskip-2pt
\ligne{\uu{\hskip 34pt C\hskip 4pt s$_1$}\hfill}
}}
}      
          {{\bf sel:}\hskip 5pt \uu C, 5$_1$; \uu D, 5$_2$} {{\bf con:} \uu s, 0} {}

\advance \regle by 6
\vskip 6pt
\novrangee {\raise 7pt\hbox{\vtop{\hsize=76pt\tt
\ligne{\uu{\hskip 19pt B\xxx A \hskip 22pt 2\xxx 1\xxx o$_r$}\hfill}
\vskip-6pt
\iii {B}{WWBBBBBBWWB}{B}
\vskip-2pt
\ligne{\uu{\hskip 20pt s\hskip 32.5pt t}\hfill}
}}
}      
          {{\bf sel:}\hskip 5pt \uu D, 5$_1$} {{\bf sen:}\hskip 5pt \uu C, 0} {}

\advance \regle by 2
\vskip 6pt
\novrangee {\raise 7pt\hbox{\vtop{\hsize=76pt\tt
\ligne{\uu{\hskip 19pt B\xxx A \hskip 22pt 2\xxx 1\xxx o$_r$}\hfill}
\vskip-6pt
\iii {W}{WWBBBBBBWWB}{W}
\vskip-2pt
\ligne{\uu{\hskip 20pt s\hskip 32.5pt t}\hfill}
}}
}      
          {{\bf sel:}\hskip 5pt \uu D, 4$_2$} {{\bf sen:}\hskip 5pt \uu C, 0} {}

\advance \regle by 14
\vskip 6pt
\novrangee {\raise 7pt\hbox{\vtop{\hsize=76pt\tt
\ligne{\uu{\hskip 28pt v\xxx E\xxx C\xxx A}\hfill}
\vskip-6pt
\iii {W}{BBBWBWBWWWW}{W} 
\vskip-2pt
\ligne{\uu{\hskip 19.5pt C\hskip 9pt o\hskip 4.5pt i}\hfill}
}}
}      
          {{\bf sel:}\hskip 5pt \uu o${_\ell}$, 6$_1$} {{\bf con:}\hskip 5pt \uu t, 0} {}
}

\end{document}